\documentclass[aps,prb,twocolumn,superscriptaddress,showpacs]{revtex4-1} 

\usepackage{amssymb}
\usepackage{graphicx}
\usepackage{amsmath}
\usepackage{dsfont}
\usepackage{bbm}
\begin{document}
\title{Interplay between edge states and simple bulk defects in graphene nanoribbons}
\author{L. Bilteanu}
\affiliation{Laboratoire de Physique des Solides, CNRS UMR-8502, Universit\'e Paris Sud, 91405 Orsay Cedex, France}
\author{C. Dutreix}
\affiliation{Laboratoire de Physique des Solides, CNRS UMR-8502, Universit\'e Paris Sud, 91405 Orsay Cedex, France}
\author{A. Jagannathan}
\affiliation{Laboratoire de Physique des Solides, CNRS UMR-8502, Universit\'e Paris Sud, 91405 Orsay Cedex, France}
\author{C. Bena}
\affiliation{Laboratoire de Physique des Solides, CNRS UMR-8502, Universit\'e Paris Sud, 91405 Orsay Cedex, France}
\affiliation{Institute de Physique Th\'eorique, CEA/Saclay, Orme des Merisiers, 91190 Gif-sur-Yvette Cedex, France}

\date{\today}

\begin{abstract} We study the interplay between the edge states and a single impurity in a zigzag graphene nanoribbon. We use tight-binding exact diagonalization techniques, as well as density functional theory calculations to obtain the eigenvalue spectrum, the eigenfunctions, as well the dependence of  the local density of states (LDOS) on energy and position. We note that roughly half of the unperturbed eigenstates in the spectrum of the finite-size ribbon hybridize with the impurity state, and the corresponding eigenvalues are shifted with respect to their unperturbed values. The maximum shift and hybridization  occur for a state whose energy is inverse proportional to the impurity potential; this energy is that of the impurity peak in the DOS spectrum. We find that the interference between the impurity and the edge gives rise to peculiar modifications of the LDOS of the nanoribbon, in particular to oscillations of the edge LDOS. These effects depend on the size of the system, and decay with the distance between the edge and the impurity. 

\end{abstract}
\maketitle

\section{Introduction }
The impact of chemical impurities on graphene electronic, optical and mechanical properties has been studied extensively, both theoretically \cite{bib1} and experimentally \cite{bib20,bib21}. The effect of impurities has been found to be very important, for example a chemical functionalization of the dangling bonds on the edges, or the chemical bonding of add-atoms to some of the bulk carbon atoms may allow to change a semi-metallic or metallic graphene sample into an insulating one \cite{bib3}. Moreover, scanning tunneling microscopy (STM) images performed on disordered samples show large-scale effects around the defect centers \cite{bib4}, associated with a perturbation of the conductance of these samples \cite{bib5}. 

While the majority of theoretical works investigating the effects of disorder on the electronic properties of graphene focus on transport \cite{transport,review},  there have been quite a number of theoretical studies focusing on the effect of disorder on the local density of states \cite{stmtheory1,stmtheory2,stmtheoryothers}. Most of these works address either the effect of a single impurity in an infinite system \cite{stmtheory1}, or the perturbations of the edge states when the impurities are localized directly on the edges \cite{stmtheory2}. Here we investigate the interplay between the edge states and the impurity states. Thus we study the perturbation of the edge states in finite size systems (e.g. graphene nanoribbons) due to impurities which are situated in the bulk. The techniques that we use are tight-binding (TB) calculations, the density functional theory (DFT) , as well as the T-matrix approximation. These calculations allow us to obtain the dependence of the local density of states (LDOS) as a function of energy and position, as well as the position dependence of the eigenfunctions of the system.



We  find that, in the presence of the impurity, the zero energy edge state wavefunctions are modified, for example their periodicity along the direction parallel to the edge of the ribbon changes. This is responsible for small oscillations of the LDOS along the edge, which decay with the distance between the impurity and the edge. Also, about half the finite energy eigenvalues are shifted, and the corresponding eigenfunctions hybridize with the impurity state. The largest hybridization and eigenvalue shift occurs for a certain state which we denote impurity state; the wavefunction corresponding to this state has maximal amplitude close to the impurity position.

Our analysis indicates that a disordered graphene nanoribbon (GNR) is a very complex system, for which the effect of the impurities cannot be treated perturbatively, nor independently from the rest of the system, and for which there is a strong interdependence between the edge states and the impurity states. Our observation may have important consequences for the analysis of transport in GNRs: it has been shown \cite{bib6} that the conduction in GNRs occurs via the edge states, and thus bulk impurities may have a strong effect on the edge transport.

The rest of this paper is organized as follows: in the next section we describe the density-functional theory (DFT), the tight-binding (TB) and the T-matrix methods. Section III  is dedicated to a discussion of the results. We present the conclusions in section IV.

\section{Model and computational techniques}

\begin{figure}[t]
\centering
\includegraphics[width=7.5cm]{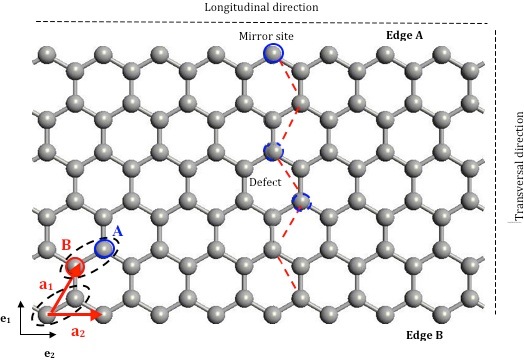}
\caption{Characteristics of a zz-GNR (exemplified for n = 3 and m = 8) : unit vectors, unit cell, sublattices, directions and mirror site definitions.}
\label{fig1}
\end{figure}

We consider a simplified system of a GNR with perfect zigzag (zz) edges and a single localized bulk defect (see Fig.~\ref{fig1}). The effects of such defect are largest on the edge containing sites belonging to the same sublattice as the defect site.  The positions of atoms in the zz-GNR are described by the units vectors $a_1$ and $a_2$ (see Figure 1) with $\|\bf{a_1}\| = \|\bf{a_2}\| = 2.461$ \r{A}, $\bf{a_1} $ = $(2.1313, 1.2305)$ \r{A} and $\bf{a_2} $ = $ (0.0, 2.461)$ \r{A}. The graphene lattice is formed by two (A and B) sublattices. We will henceforth take the defect to be located on an A atom. We refer to the direction parallel to the edge as longitudinal, and to the direction perpendicular to the edge as transversal.  When analyzing the behavior of the LDOS along the transversal direction we will consider all the A atoms lying in a zigzag path (the red dashed line in Fig.~\ref{fig1}).  We define as the impurity `mirror' site the atom closest to the projection of the impurity on the A edge; depending on the distance between the impurity and the edge, this site can fall directly above the impurity, or be displaced laterally by one atom (see Fig.~\ref{fig1}); in the latter case there exist actually two equivalent mirror sites.


The zz-GNR is characterized by two chiral numbers $(n, m)$. The first chiral number, $n$, is related to the width of the ribbon:\\
\begin{equation}
w\left( n \right) =\frac{2}{3}\left( 3n-1 \right)\left\| {{\mathbf{a}}_{1}}\cdot{{\mathbf{e}}_{1}} \right\|=\frac{2 \left( 3n-1 \right){{w}_{0}}}{3}
\end{equation}
where we have defined the following width constant (that will be used in the next section) : 
\begin{equation}
    {{w}_{0}}=\left\| {{\mathbf{a}}_{1}}\cdot {{\mathbf{e}}_{1}} \right\|=2.1313 \AA
\end{equation}
The second chiral number, $m$, is the number of repetitions along the longitudinal direction and hence it is related to the length of the zz-GNR by
\begin{equation}
L=m \left\| {{\mathbf{a}}_{2}}\cdot {{\mathbf{e}}_{2}} \right\|
\end{equation}
The number of atoms in a zz-ribbon is dependent on both $n$ and $m$ :
\begin{equation}
N = 4 n \cdot m
\end{equation}
We present the corresponding characteristics for the set of ribbons that we study in this paper in Table 1.  While we use a fixed ribbon length, we have checked that modifying this length does not qualitatively change our results.

\begin{table}[ht]
\caption{Ribbon characteristics (chiral numbers, width, length, number of atoms), as well as the method of choice for each GNR considered.} 
\centering  
\begin{tabular}{c c c c c c c} 
\hline\hline                        
n & m & w(\r{A}) & L (\r{A}) & N & w/L & Method  \\ 
\hline\hline                  
5  & 20 &  19.89 & 49.22 &  400 & 0.40 &  TB  \\ 
6  & 20 &  24.15 & 49.22 &  480 & 0.49 &  TB  \\
7  & 20 &  28.42 & 49.22 &  560 & 0.58 &  TB  \\
8  & 20 &  32.68 & 49.22 &  640 & 0.66 &  TB  \\
9  & 20 &  36.94 & 49.22 &  720 & 0.75 &  TB  \\
10 & 20 &  41.21 & 49.22 &  800 & 0.84 &  TB  \\
15 & 20 &  65.52 & 49.22 &  1200 & 1.33 &  TB  \\
20 & 20 &  83.83 & 49.22 &  1600 & 1.70 &  TB  \\
25 & 20 & 105.14 & 49.22 & 2000 & 2.13 &  TB  \\
5  & 10 &  19.89 & 24.61 &  200 & 0.81 & DFT  \\
6  & 10 &  24.15 & 24.61&  240 & 0.98 & DFT  \\
7  & 10 &  28.42 & 24.61 &  280 & 1.15 & DFT  \\
8  & 10 &  32.68 & 24.61 &  320 & 1.33 & DFT  \\
9  & 10 &  36.94 & 24.61 &  360 & 1.50 & DFT  \\
10 & 10 &  41.21 & 24.61 &  400 & 1.67 & DFT  \\
\hline 
\end{tabular}
\end{table}

\subsection{Tight-binding method}
For the tight-binding calculations we have performed an exact diagonalization of the one-orbital Hamiltonian:
\begin{equation}
	\widehat{H}=\sum\limits_{i,j}{{{t}_{ij}}\left| i \right\rangle \left\langle  j \right|}+\sum\limits_{i}{{{V}_{i}}\left| i \right\rangle \left\langle  j \right|}
\label{ham}
\end{equation}
where $t_{ij}$ is the hopping parameter between the $p_z$ orbitals $\left| i \right\rangle $  and $\left| j \right\rangle $. The second term is modeling the presence of defects on a finite number of graphene sites, via the introduction of the on-site potentials $V_i$. In this paper we consider the first-order, nearest-neighbor approximation ($t_{ij}$ is $-1$ when $i$ and $j$ are nearest-neighbor sites and zero otherwise). Taking into account realistic values for higher-order hopping processes \cite{t2t3} does not change qualitatively the results presented here, though unrealistically large values for these hopping terms may strongly affect the formation of edge states, as well the interaction between the edges and the impurity. These effects will be addressed elsewhere \cite{t3}. We assume semi-periodic boundary conditions (periodic in the longitudinal direction and open in the transversal direction). 

When taking into account the effect of a single defect localized on the site $d$, the  Hamiltonian (\ref{ham}) can be written as: 
\begin{equation}
\widehat{H}=-\sum\limits_{\left\langle i,j \right\rangle }{\left| i \right\rangle \left\langle  j \right|}+{{V}}\left| d \right\rangle \left\langle  d \right|
\label{ham2}
\end{equation}
where the summation  is performed over the nearest-neighbor sites.  The eigenfunctions of (\ref{ham2}) can be written as a linear combination of individual orbitals:  

\begin{eqnarray}
\left| k \right\rangle =\sum\limits_{i}{{{c}_{ki}}\left| i \right\rangle},\\
{{c}_{ki}}=\left\langle  i | k \right\rangle
\end{eqnarray}
where $\left| k \right\rangle$ is the level index corresponding to the energy $E_k$.


The Hamiltonian (\ref{ham2}) can be diagonalized, and the corresponding eigenvalues $E_k$ and eigenvectors  (respectively $c_{ki}$) can be obtained numerically. One defines the total density of states:
\begin{equation}
\rho\left( E \right)=\sum\limits_{k}{\delta \left( E-{{E}_{k}} \right)},
\label{ddos}
\end{equation}
where $\delta \left( E \right)$ is the Dirac delta function. While the DOS spectrum defined by Eq.~(\ref{ddos}) is discrete, for realistic systems a phenomenological rounding of the delta function needs to be introduced:
\begin{equation}
\rho\left( E \right)=\sum\limits_{k}{f\left( E-{{E}_{k}} \right)}
\end{equation} 
where we take $f$ to be a Lorentzian with a width of $0.05$.

Since we are interested not only in the average density of states, but also in the dependence of the density of states with position, we define also a local density of states (LDOS): 
\begin{equation}
{{\rho}}\left(\vec{r}_i, E \right)=\sum\limits_{k}{{{\left| {{c}_{ki}} \right|}^{2}}f\left( E-{{E}_{k}} \right)}
\end{equation}
 
\subsection{Density functional theory}
All DFT calculations presented in this paper have been performed using the QuantumWise code which is a DFT implementation using pseudopotentials and numerical atomic orbitals (NAO). For all our calculations we have employed the single-zeta polarized basis set and the pseudopotentials provided by the package. In order to find out the defect structure we have performed full relaxation calculations: the GNR was allowed to relax in all directions, and the dynamic convergence criterion for the relaxation calculation was a maximal force of $0.05 eV/\r{A}$, with the energy convergence condition of each self-consistent calculation of $10^{-5} eV$. The carbon dangling bonds on the edges have been terminated with $H$ atoms, since the dangling bonds are not well treated in the DFT framework. We have checked that the standard GNR electronic structure (band structure and total density of states) is retrieved even in the presence of hydrogenation.

We have investigated the effect of a vacancy inside ribbons of different widths ($n = 5 - 10$), but having the same length (see Table 1).  For the largest ribbon analyzed, we have investigated the effects of a vacancy on the edge when the position of the vacancy is modified.  The different sizes of the nanoribbons used in the calculations imposed different choices for the $k$-sampling of the $k$-space: for greater widths we have employed more $k$-points in the Monkhorst-Pack scheme ($1\times 3\times 3$ to $1 \times 5\times 5$). The sampling has been chosen by performing accuracy tests for the calculation of the formation energy of a vacancy inside the graphene nanoribbon. Inside a graphene sheet a fully relaxed vacancy can adopt two configurations \cite{bib10}: a configuration in which all the atoms are in the graphitic plan (planar configuration) and a second in which one of the atoms is slightly out of the graphitic plan (off-plane configuration). The second configuration is more stable than the first one. The converged values for the formation energy of a monovacancy in the first configuration inside different nanoribbons were in excellent agreement with values obtained in previous works \cite{bib10}. These values were not sensitive to modeling the vacancy as a ghost atom.

QuantumWise packages provide visual tools to access directly to a broad range of numerical outputs among which the most important for the purpose of this investigation are the total DOS, the dependence of the DOS with energy at a given site, and the dependence of the DOS as a function of position at a given energy. 

\subsection{T-Matrix approximation}
The LDOS perturbations in a finite-size system in the presence of disorder obtained numerically via DFT and TB can be compared to those corresponding to an infinite system. For such a system, the effect of an impurity on the LDOS can be evaluated using the T-matrix approximation\cite{stmtheory1}. This method consists in treating the effects of the defect perturbatively, and using diagrammatic Green's function techniques. Since this technique has been extensively used in the past \cite{stmtheory1}, the results obtained via this technique will not be the central focus of this paper, but we will use it rather as a cross-check for the DFT and TB calculations, as well as to compare the LDOS in the presence of a single impurity for the infinite and finite-size systems.

\section{Results}
In this section we present the results obtained through the TB (section III A), DFT (section III B) and T-matrix (section III C) techniques. 
\subsection{Tight-binding calculations}
Here we focus on the results obtained via an exact diagonalization of the TB Hamiltonian. We first classify the system eigenfunctions for a defect-free (df) and a defect-containing (dc) GNR. Secondly we present the variation of the LDOS at zero energy  along an axis parallel and respectively perpendicular to the edge of the GNR. Finally we analyze the two-dimensional profile of the LDOS at various energies for df-GNRs and dc-GNRs. 

\subsubsection{The eigenvalue spectrum and the corresponding eigenfunctions}
As it has been previously shown \cite{dresselhaus,bib16,bib17}, the DOS spectrum of zz-GNR exhibits a central peak. If in the tight-binding model we only take into account the nearest-neighbor hopping processes, this peak is located at zero energy, and the spectrum is symmetric with respect to $E = 0$. We will focus mostly on this situation in what follows. Different values of the higher-order hopping terms will be considered elsewhere \cite{t2t3}. Typical DOS spectra for a df-GNR and dc-GNR described by a tight-binding hopping Hamiltonian with semiperiodic boundary conditions (periodic along the longitudinal direction and open along the transversal one) are presented in Fig.~\ref{ddos}; note the central peak corresponding to the edge states, as well as a smaller low-energy peak corresponding to the impurity state \cite{stmtheory1}.

\begin{figure}[t]
\centering
\includegraphics[width=7cm]{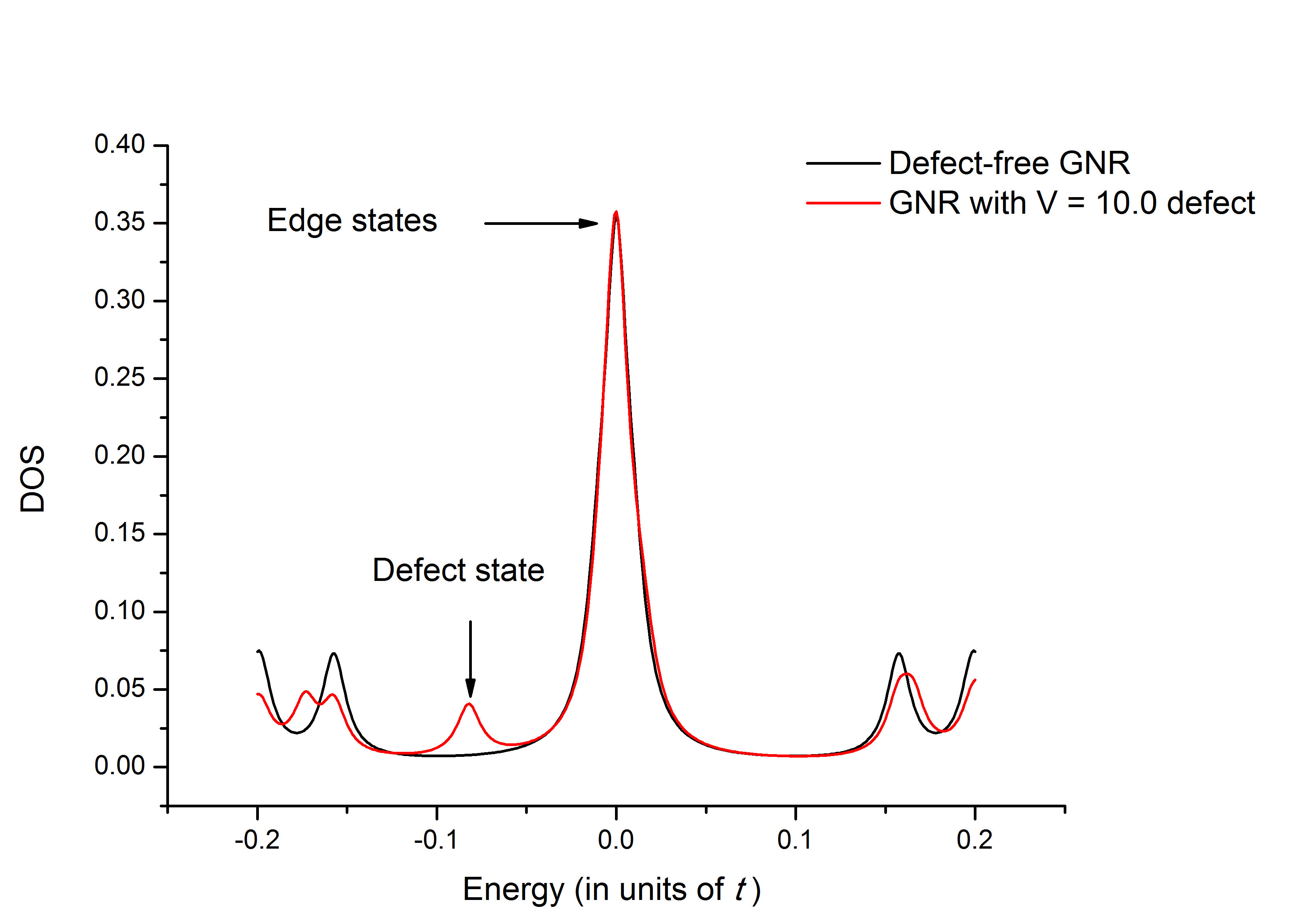}
\label{dos}
\caption{TB calculated DOS in a df-GNR and dc-GNR ($n=15$). Note the central peak corresponding to the edge states, as well as the small low-energy peak corresponding to the impurity state.}	
\end{figure}

\begin{figure}[t]
\centering
\includegraphics[width=8cm]{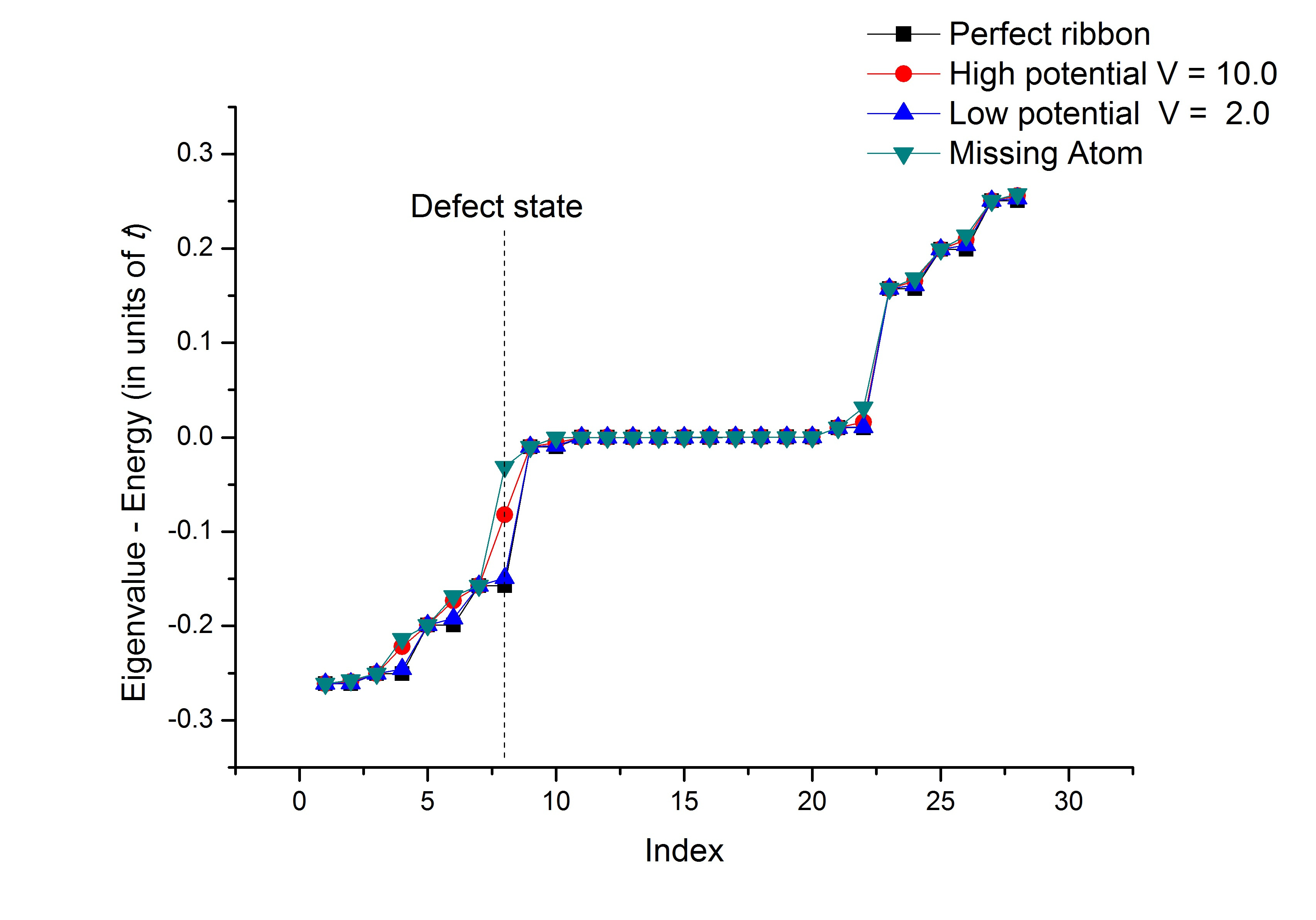}
\caption{The lowest-energy eigenstates for a $n=15$ GNR spectrum for an unperturbed system, as well as for different values of the impurity potential.}
\label{eigenvalues}
\end{figure}

\begin{widetext}

\begin{figure}[t]
\centering
\includegraphics[width=4cm]{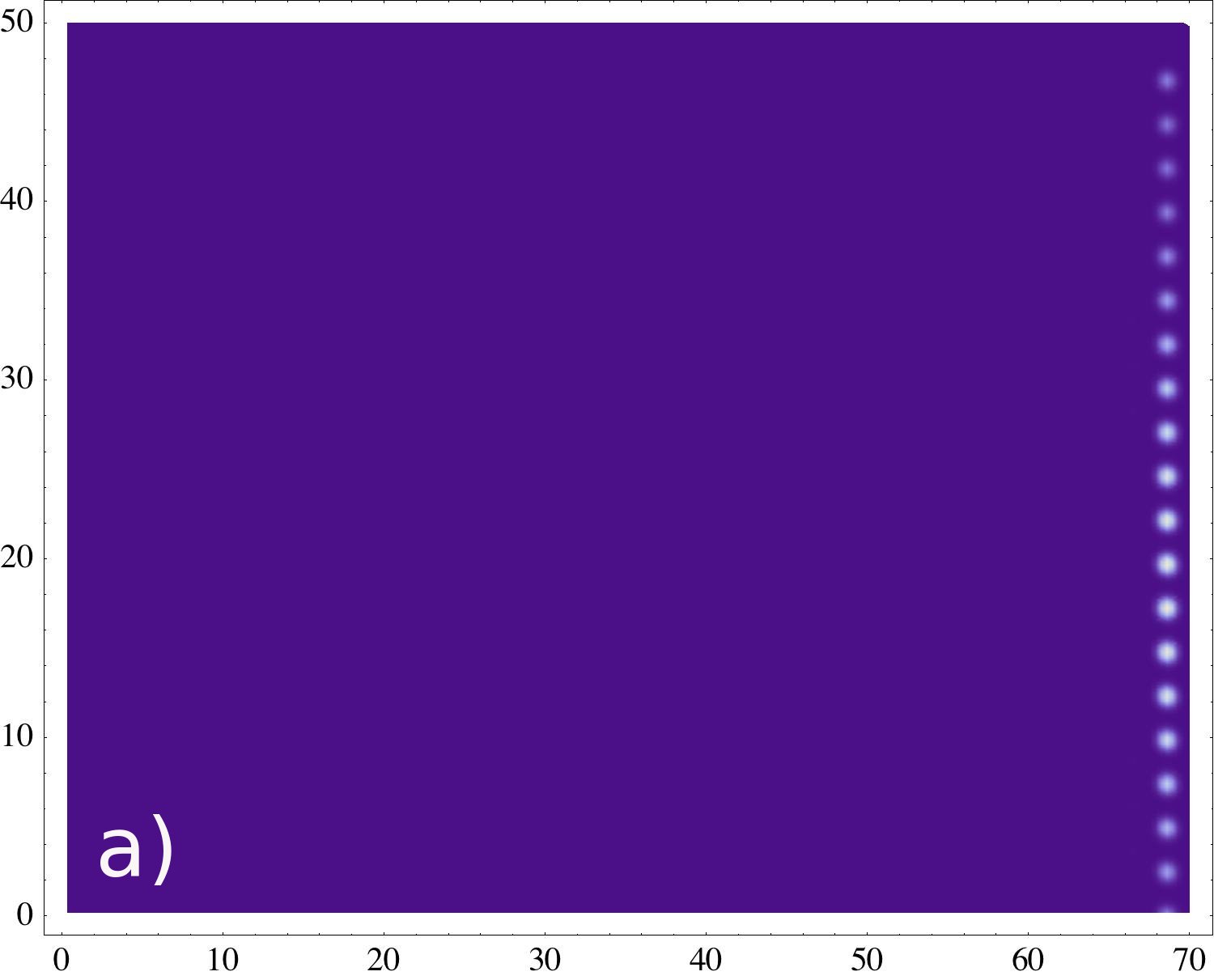}
\includegraphics[width=4cm]{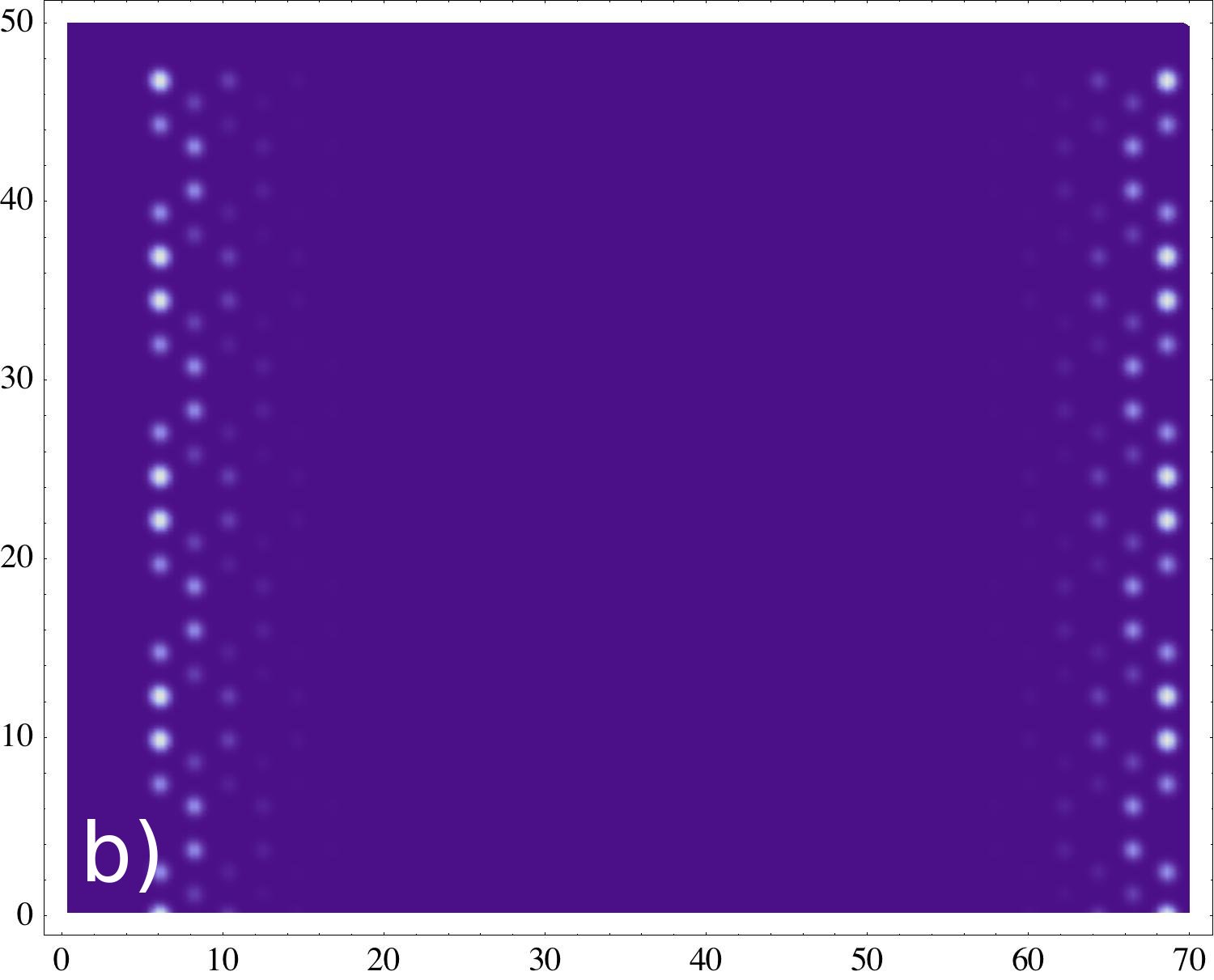}
\includegraphics[width=4cm]{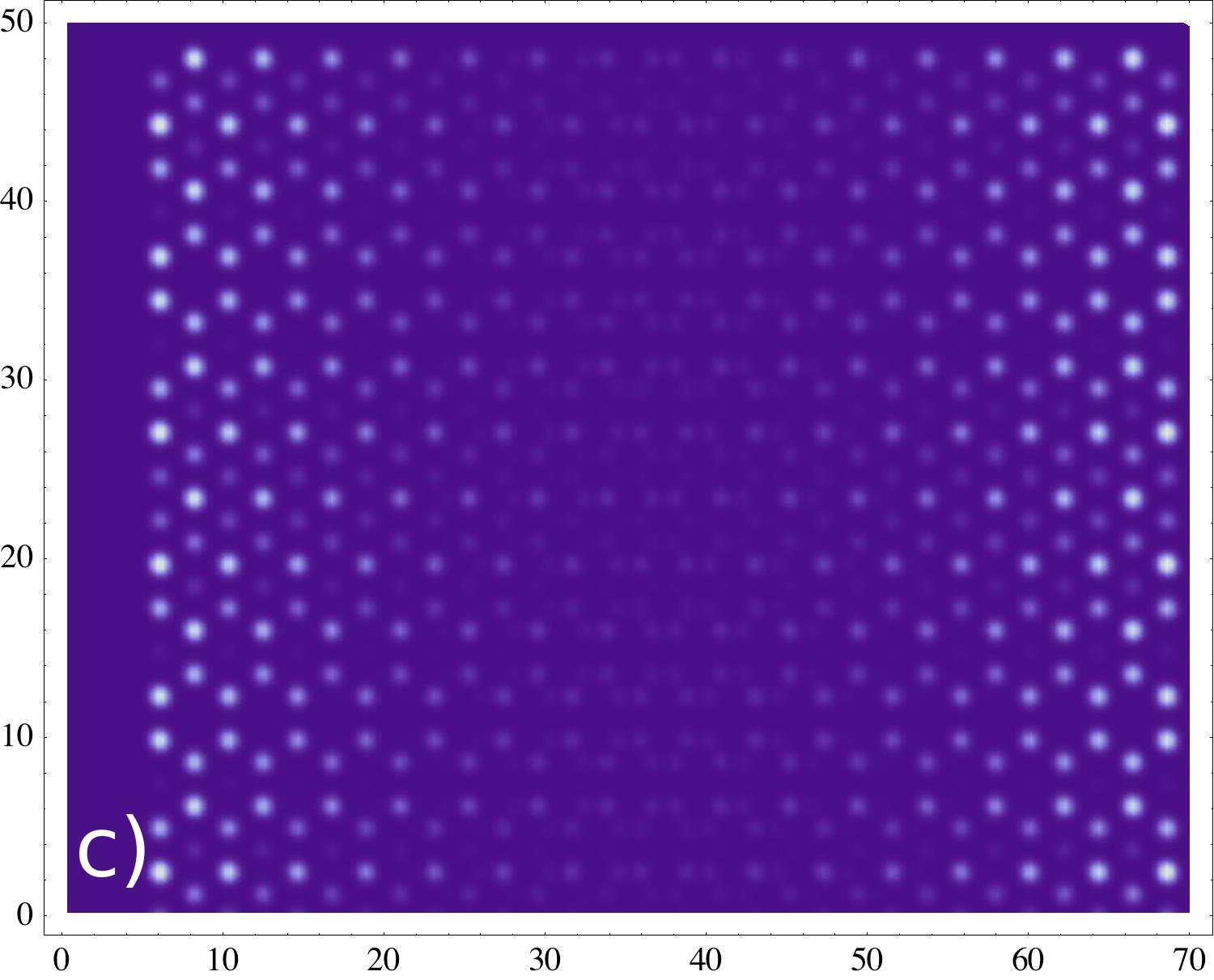}
\includegraphics[width=4cm]{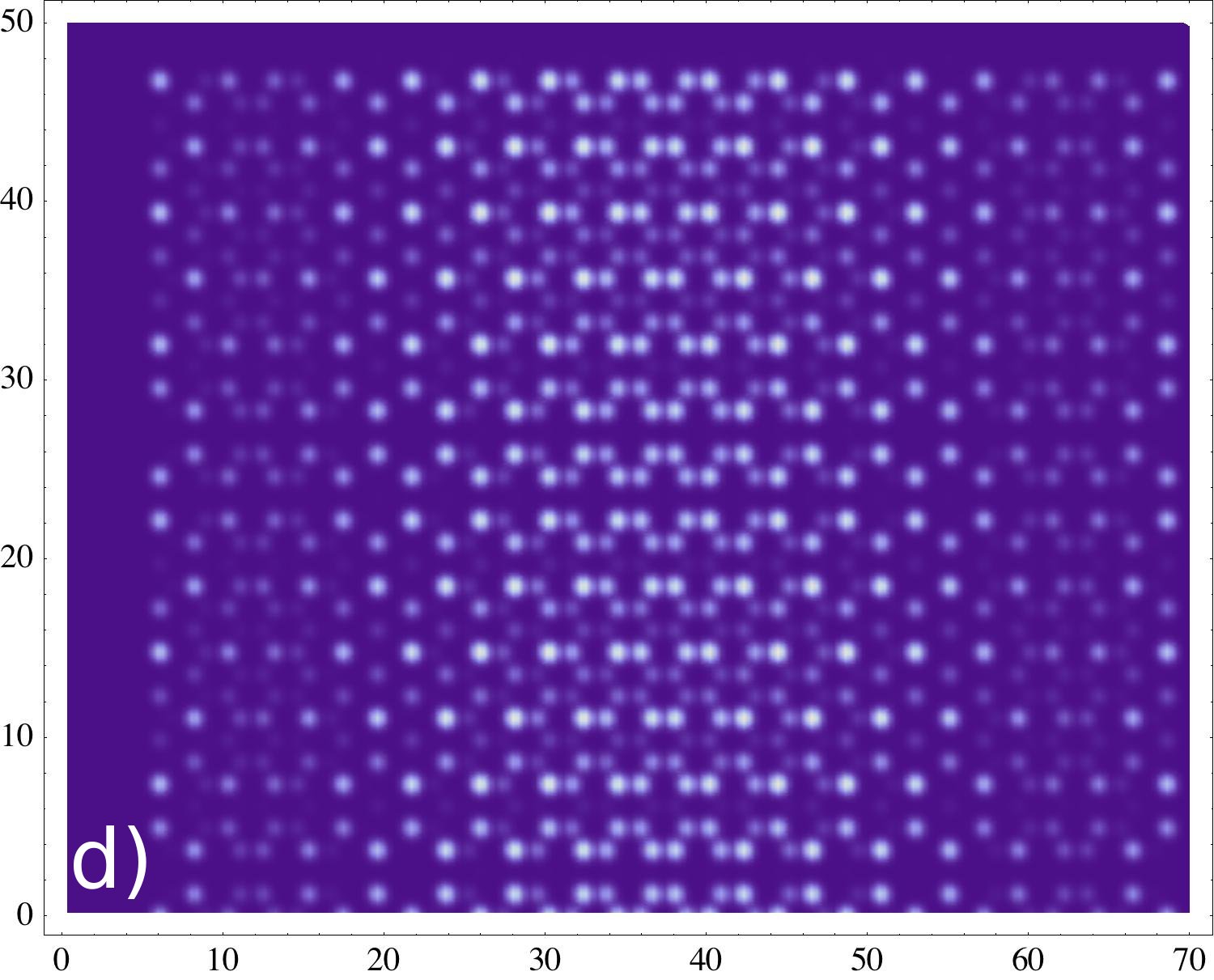}
\includegraphics[width=4cm]{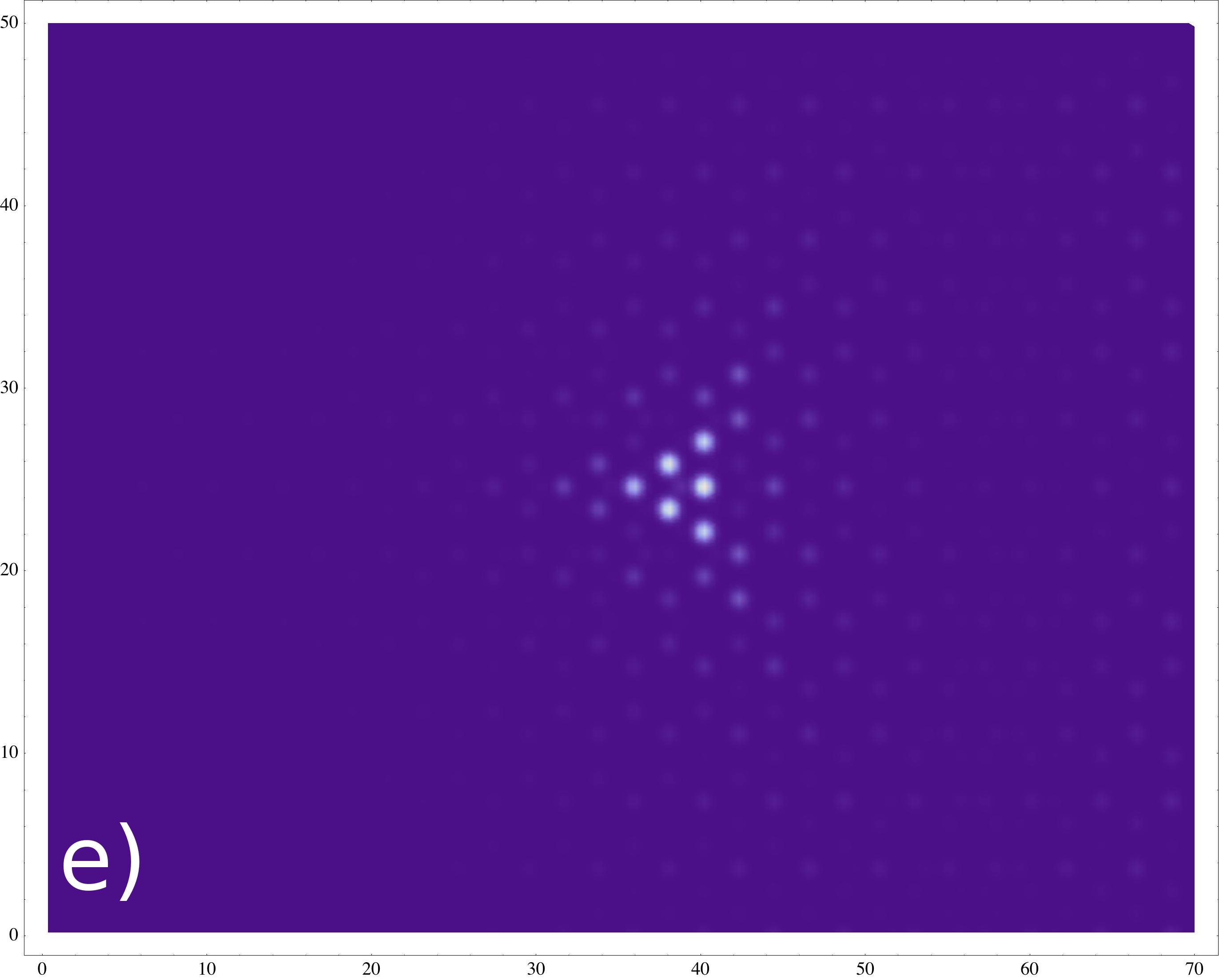}
\includegraphics[width=4cm]{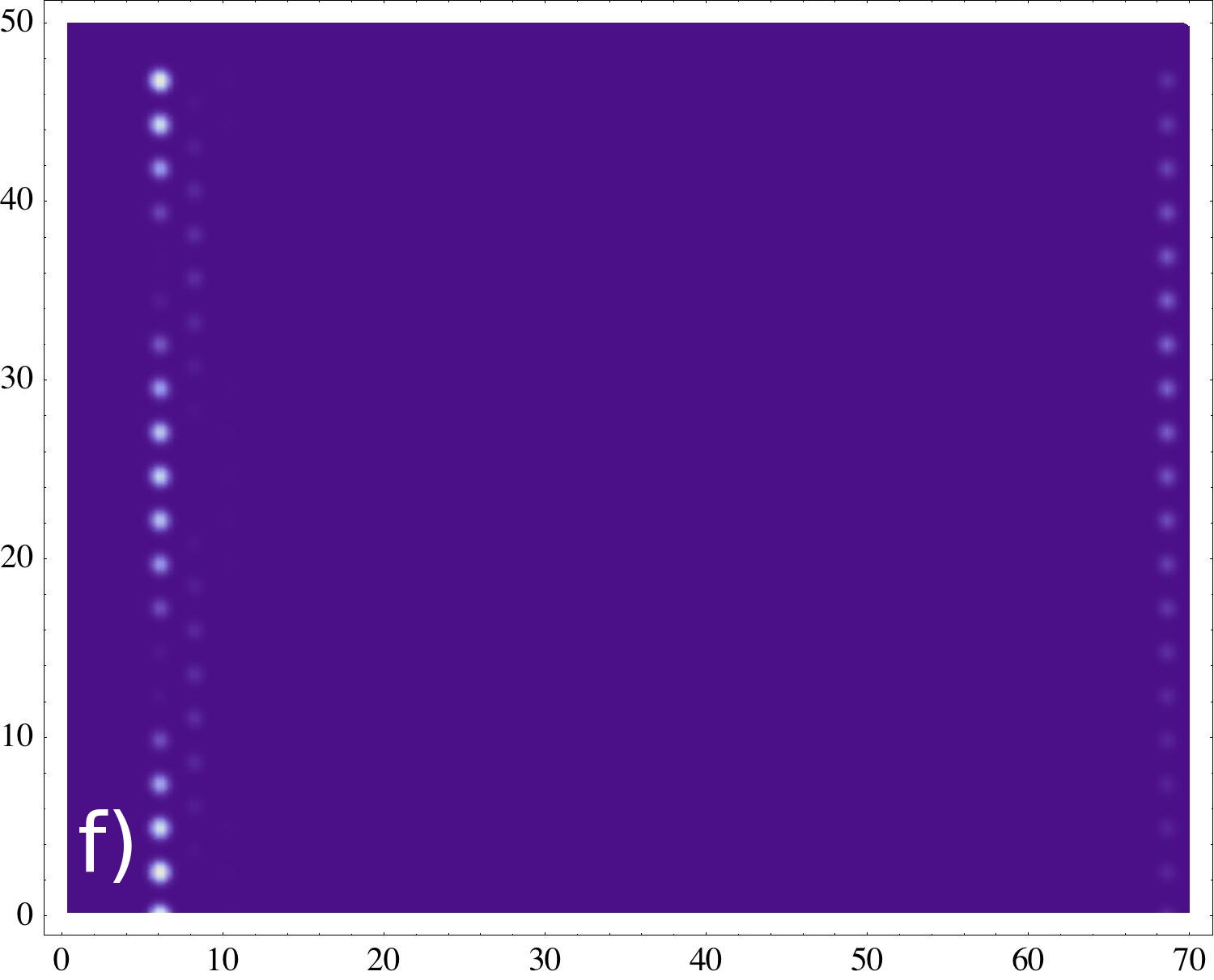}
\includegraphics[width=4cm]{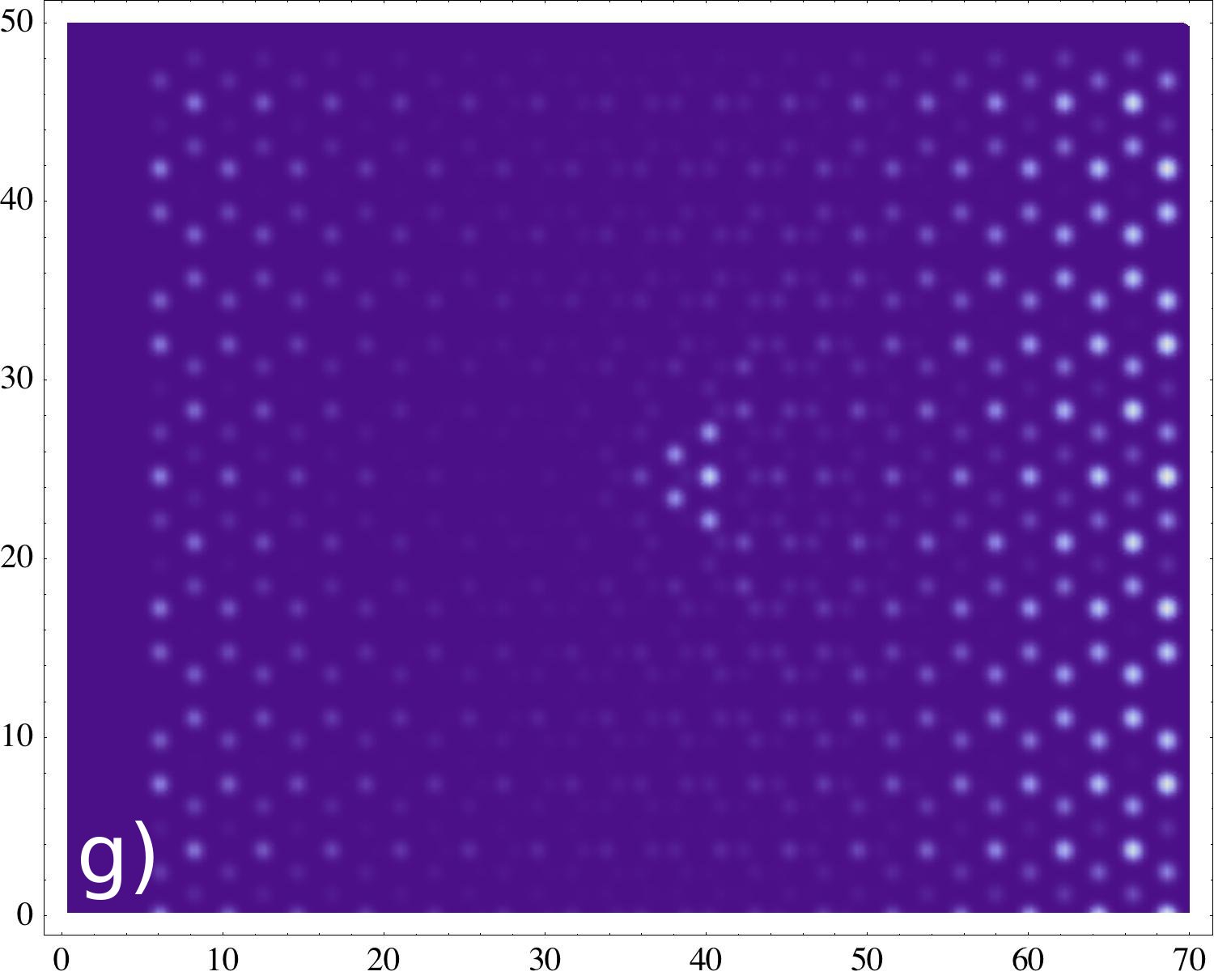}
\includegraphics[width=4cm]{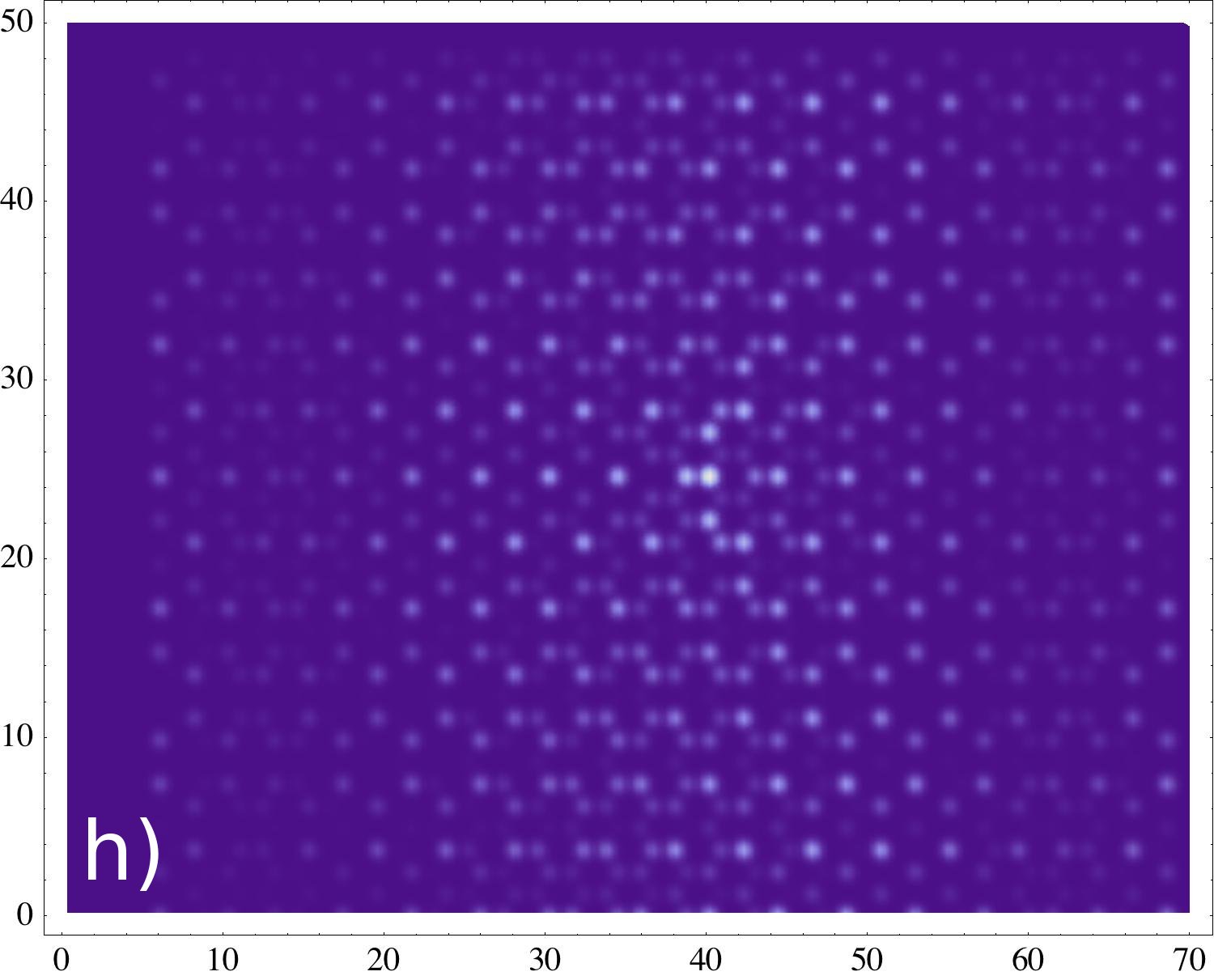}
\caption{Wavefunction typology as resulting from the TB analysis: (a) Localized zero-energy edge states in a $n=15$ df-GNR; (b) Low-energy edge states in df-GNR - note the non-zero but small bulk extension; (c) Bulk-edge states in df-GNR - the bulk extension is significative; (d) Bulk states in df-GNR;  (e) Impurity state for a dc-GNR ($V = 10$);  (f) Perturbed edge states in a dc-GNR, note that such states do not hybridize with the impurity, but their edge periodicity  is modified; (g) Hybrid bulk-edge impurity state for a dc-GNR ($V = 10$);  (h) Hybrid bulk-impurity state for a dc-GNR ($V = 2$);}	
\label{fig7}
\end{figure}

\end{widetext}

An analysis of the eigenvalues of the tight-binding Hamiltonian reveals that the central peak in the DOS corresponds to several low-energy states (see Fig.~\ref{eigenvalues}). 
This analysis also reveals that, in the presence of a defect, the eigenvalue spectrum is modified, such that approximatively half of the finite energy eigenvalues are shifted. This shift, which decreases with the value of the impurity potential,  is maximum for a certain state which we denote as `defect' state or `impurity' state (see Fig.~\ref{eigenvalues}). The formation of this state at an energy for which there is no corresponding state in the clean ribbon is responsible for the DOS impurity peak depicted in Fig.~\ref{ddos}.  As we will show in what follows (see Fig.~\ref{fig7} e), the eigenfunction corresponding to this eigenvalue is mostly localized in the vicinity of the impurity, justifying the identification of this state as `impurity state'.

The eigenstates corresponding to the eigenvalue spectrum described above can be classified according to their energy and their position dependence (all energies are evaluated in units of the hopping parameter $t$ which we take to be equal to $1$). Thus, for a $n=15$  clean nanoribbon, the lowest energy eigenstates range in energy from  $10^{-16}$ or $10^{-15}$ (zero energy in the limit of the numerical precision), up to about $10^{-2}$. While there are always only two eigenstates localized exclusively on the edges, the number of `edge' states (states that have a maximum intensity on the edge and a non-zero decay in the bulk) varies with the size of the system. With increasing the ribbon size, the overall number of eigenvalues, and consequently the number of eigenstates composing the zero-energy peak increases (for narrow zz-GNR, $n < 5$, there are only two edge eigenstates).

The lowest-energy wavefunctions are localized on the edge  (see Fig.~\ref{fig7} a,b) and exhibit oscillations along the longitudinal direction, with a period inverse proportional to their corresponding energy. When the energy increases, besides the longitudinal oscillations, the wavefunctions also exhibit a non-zero exponential decay in the bulk, whose extent is proportional to their energy (see Fig.~\ref{fig7} b,c). This is consistent with the traditional edge-state picture for graphene \cite{dresselhaus,bib16,bib17}. The highest-energy edge states have the slowest bulk decay and very little longitudinal variation; we denote these states as bulk-edge states. Above a certain energy value (in the considered example of the order of $10^{-1}$) the character  of the wavefunctions becomes fully bulk, and the direction of their modulation changes from being parallel to the edge to being transversal  (see Fig.~\ref{fig7} d).

In the presence of the defect, the lowest-energy edge states exhibit a modified periodicity with respect to the unperturbed case (Fig.~\ref{fig7}  f), but no actual hybridization occurs between the defect and the edge states (the wavefunctions do not show an increased intensity close to the defect position). However, some of the higher-energy bulk-edge states and bulk states hybridize with the impurity state (Fig.~\ref{fig7}  g,h). More precisely, as also indicated in Fig.~\ref{eigenvalues}, we find that the two-fold degeneracy of the eigenvalue spectrum is broken by the impurity, such that, for each pair of eigenvalues, one eigenvalue and its corresponding eigenstates are unchanged, while the other eigenvalue is shifted, and the corresponding eigenstate exhibits a hybridization between a GNR wavefunction and the defect state. The defect state, for which the wave function amplitude is maximal in the vicinity of the impurity is depicted in Fig.~\ref{fig7} e).

For a high value of the impurity potential, or in the presence of a vacancy (which is equivalent to an infinite repulsive on-site potential) the eigenvalue shift is significant only for a small number of states with energies close to that of the defect state (see Fig.~\ref{eigenvalues}) which is approaching zero in this case. For an infinite system this translates into a sharp impurity peak in the DOS close to zero energy\cite{stmtheory1}. For a smaller value of $V$, the defect state occurs at higher energies, and, as one can see from Fig.~\ref{eigenvalues}, a larger number of eigenvalues are shifted by smaller amounts. This corresponds to the formation of a broader and less intense peak in the DOS of a infinite system in the presence of small $V$ impurities \cite{stmtheory1}.

\begin{figure}[t]
\centering
\includegraphics[width=8cm]{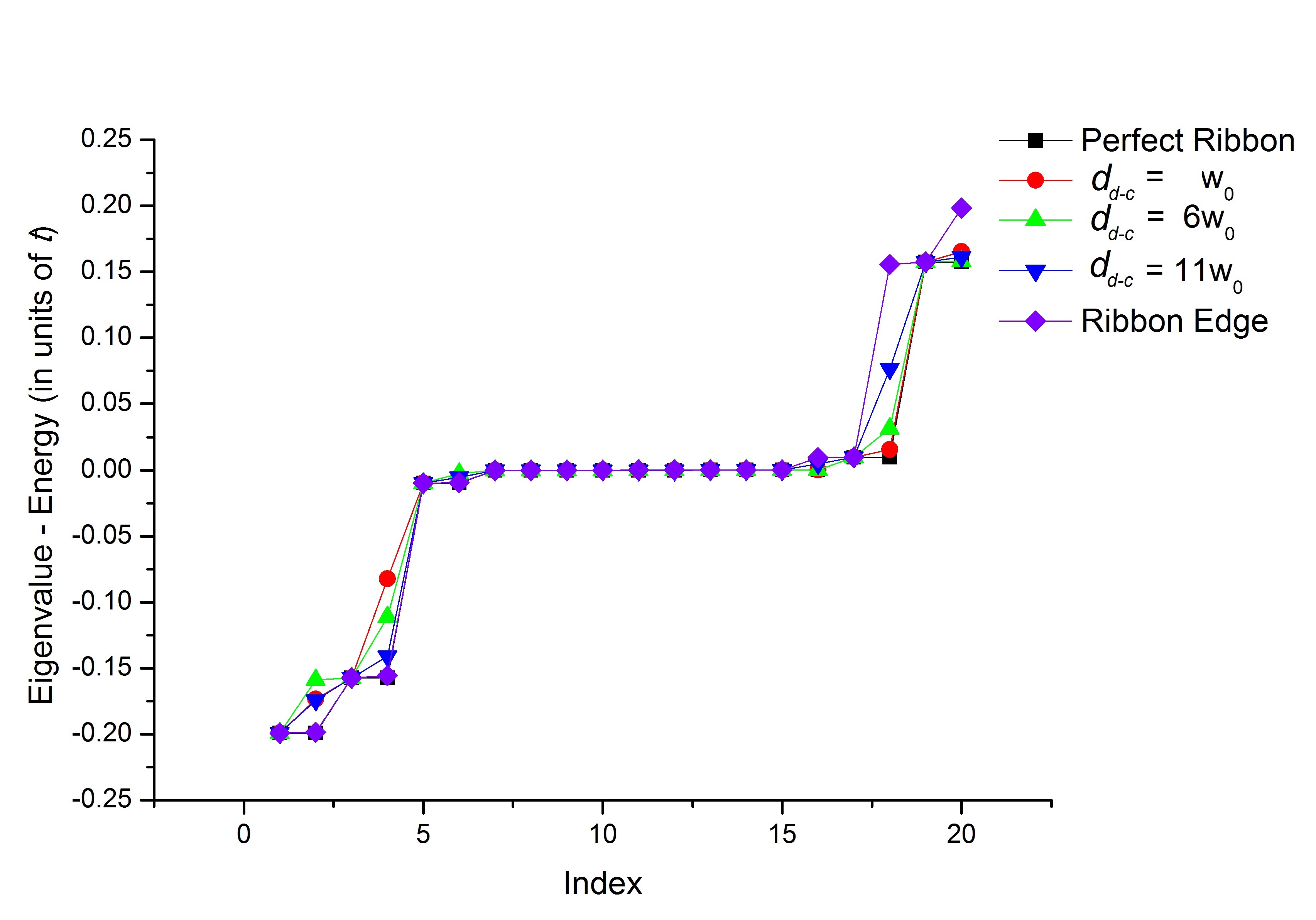}
\caption{The lowest-energy eigenvalues for a $n=15$ GNR and for different impurity positions (here $d_{d-c}$ is the distance between the impurity and the ribbons center).}
\label{eig2}
\end{figure}

We have also observed that the eigenvalue shift and the energy of the defect state depend on the distance between the edge and the impurity. In Fig.~\ref{eig2} we depict the corresponding eigenvalues when the distance between the edge and the defect is reduced.

Note that when the impurity is getting very close to the edge, the eigenstate that exhibits the largest eigenvalue shift does not correspond to the first negative-energy bulk state, but to the first positive-energy bulk state, as depicted in Fig.~\ref{eig2}.

\subsubsection{Spatial variations of the zero-energy LDOS along special directions}
\begin{figure}[t]
\centering
\includegraphics[width=8.2cm]{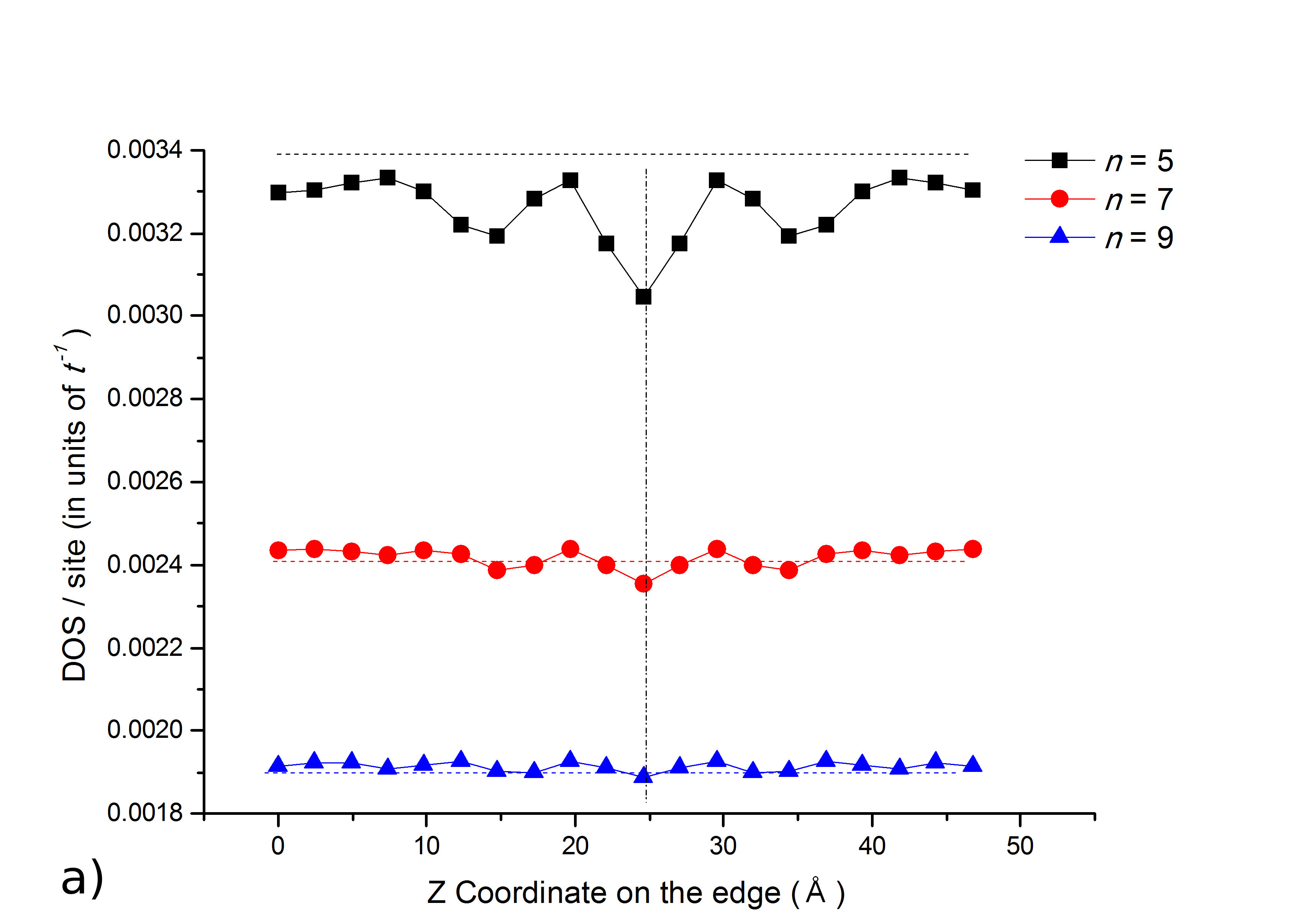}
\includegraphics[width=8.2cm]{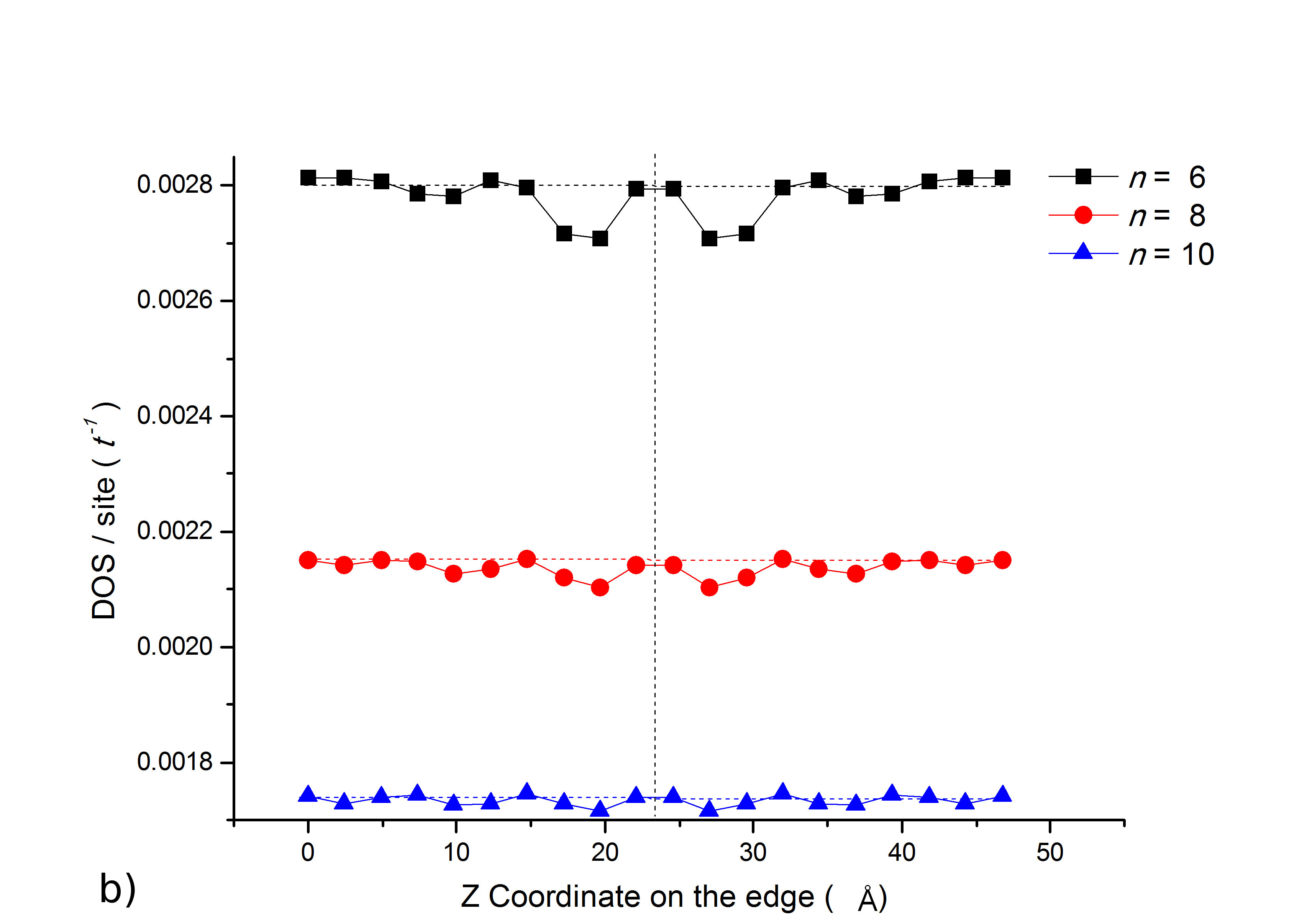}
\caption{Dependence of the LDOS on position along the edge for ribbons having different zz-GNR widths : (a) odd and (b) even $n$.}
\label{fig2}
\end{figure}

The characteristcs of the wavefunctions described in the previous section have a direct influence on the behavior of the LDOS as a function of position at different energies. In this section we analyze the dependence of the zero-energy density of states as a function of position. The effects of the impurity are most pronounced on the edge made of atoms of the same sublattice as the impurity, while the effects of the impurity on the opposite edge are much smaller. We focus on the dependence of the DOS along one of the two special directions (longitudinal and transversal)  described in Fig.~\ref{fig1}. 
\paragraph{The variation of the zero-energy LDOS along the edge}
\label{secd1}
We should first note that in order to accurately study the perturbations induced by the impurity on the edge states, one needs to focus on ribbons that are sufficiently wide ($n>5$). 
From a  TB analysis of such dc-GNRs we find that the presence of a defect perturbs the zero-energy edge density of states such that it oscillates, exhibiting minima and maxima. The magnitude of these oscillations depends on two factors: the width of the ribbon (as well as the distance from the defect site to the edge), and the value of the impurity potential. The form of the perturbation depends on the distance from the defect site to the edge: if this distance is an even multiple of $w_0$ the edge DOS exhibits a maximum on the defect mirror site, while if it is an odd multiple of $w_0$, the edge DOS exhibits a minimum (Fig.~\ref{fig2}). This is probably the effect of the positioning of the central impurity exactly on top of the mirror site for odd chiral number $n$'s, and to its lateral displacement by one atom for even $n$'s, as described in Fig.~\ref{fig1}. As expected, in all ribbons the amplitude of the edge DOS oscillations is proportional to the value of the onsite potential.

\begin{figure}[t]
\centering
\includegraphics[width=8cm]{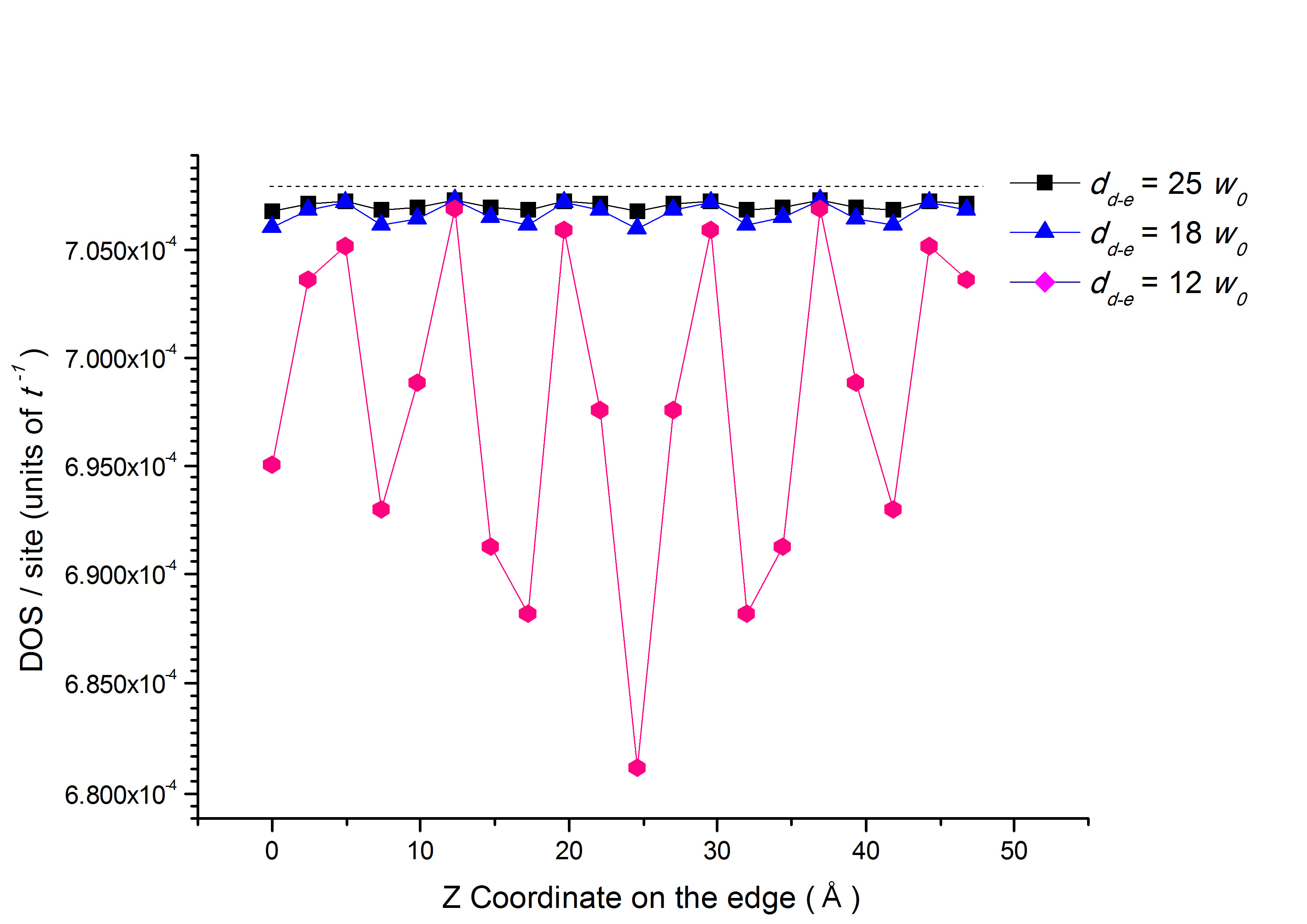}
\caption{Edge DOS oscillations for different distances between the defect site and the edge $d_{d-e}$, for a $n=25$  ribbon and a $V=10$.}
\label{fig4}
\end{figure}

The amplitude of the edge state oscillations depends also on the distance between the defect and the edge. To quantify this effect we plot the dependence of the LDOS along the edge for various distances between the edge and the defect (see Figure.~\ref{fig4}). While for defects relatively far from the edge rather uniform oscillations 
are observed, with an amplitude which is decreasing with the distance between the edge and the defect, for defects close to the edge, the amplitude of the DOS on the sites that are closer to the defect is diminishing drastically with respect to the intensity on sites further away.

\begin{figure}[t]
\centering
\includegraphics[width=8cm]{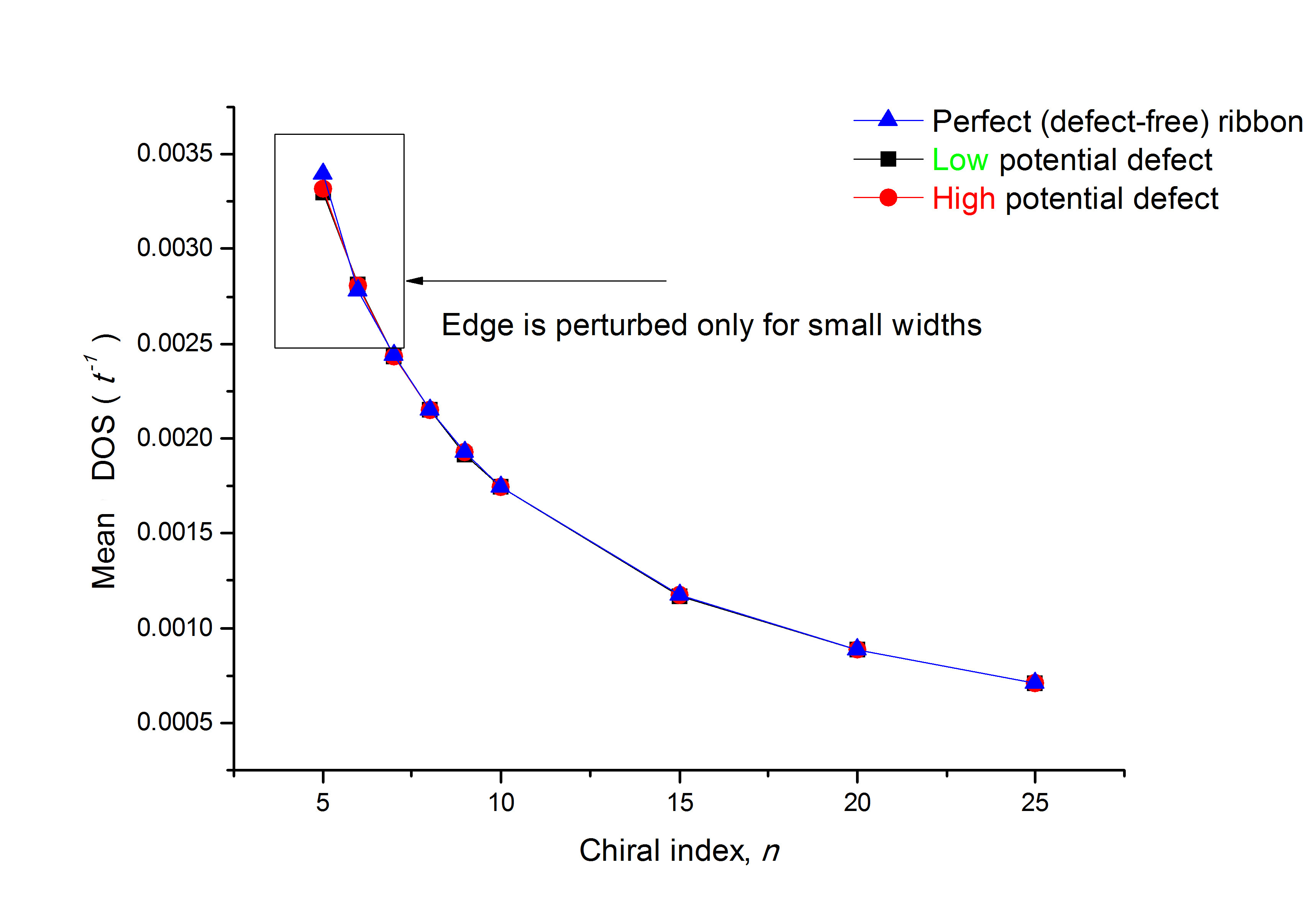}
\caption{Mean edge DOS as a function of the ribbon width in df-GNR and dc-GNR. The mean value of the edge DOS in the presence of the defect is modified only for very narrow ribbons. Note the strong decay of the mean edge DOS with the ribbon width.}
\label{fig3}
\end{figure}

The average edge DOS is also affected by the distance from the defect to the edge. While, when the defect is far away from the edge the mean edge DOS is basically unperturbed from the defect-free situation (see Fig.~\ref{fig3}), when the distance from the defect to the edge is reduced, the mean edge DOS is reduced, as the electrons are pushed away from the edge by the repulsive impurity potential. 

\begin{figure}[t]
\centering
\includegraphics[width=8cm]{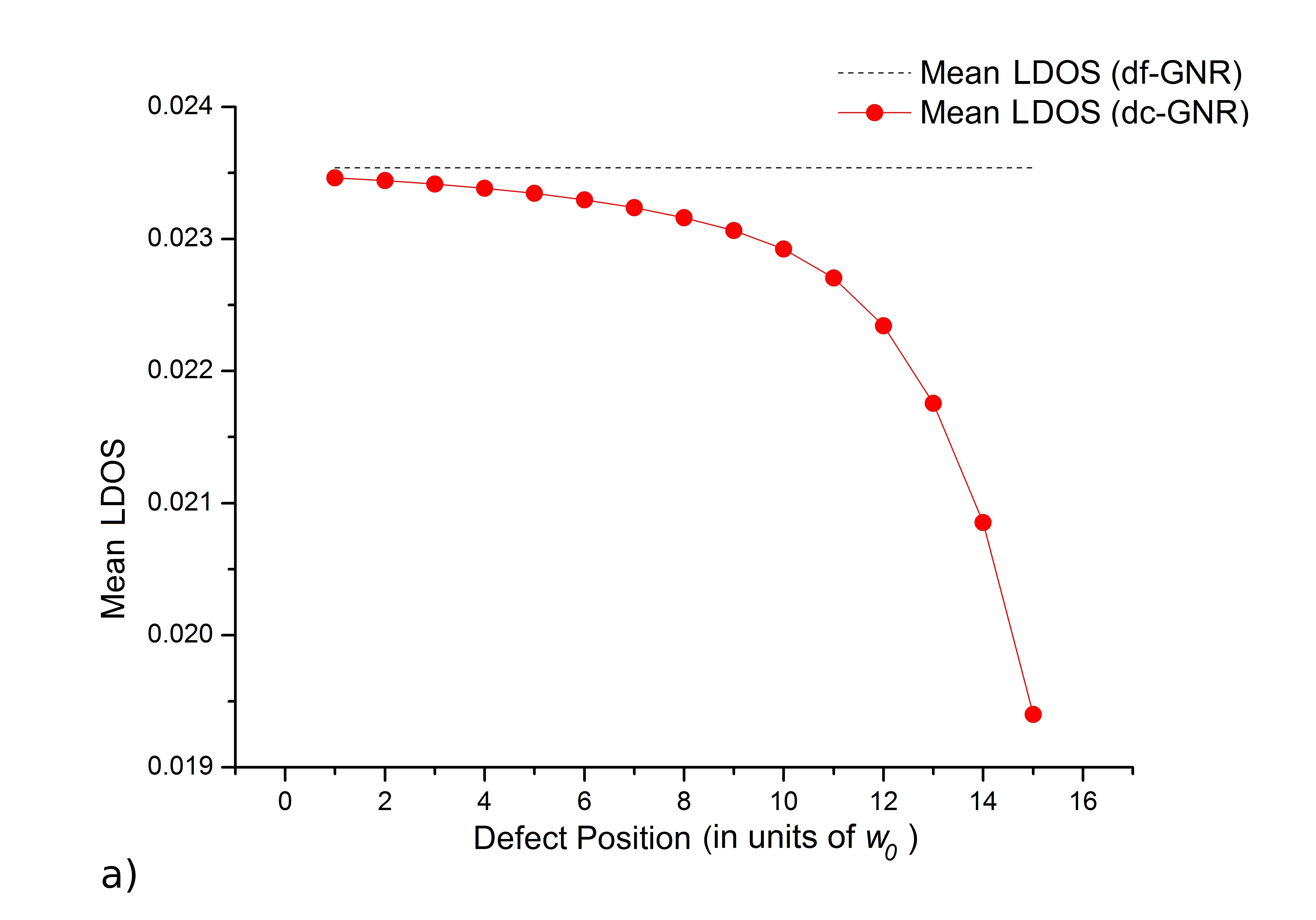}\\
\vspace{.5cm}
\includegraphics[width=6.5cm]{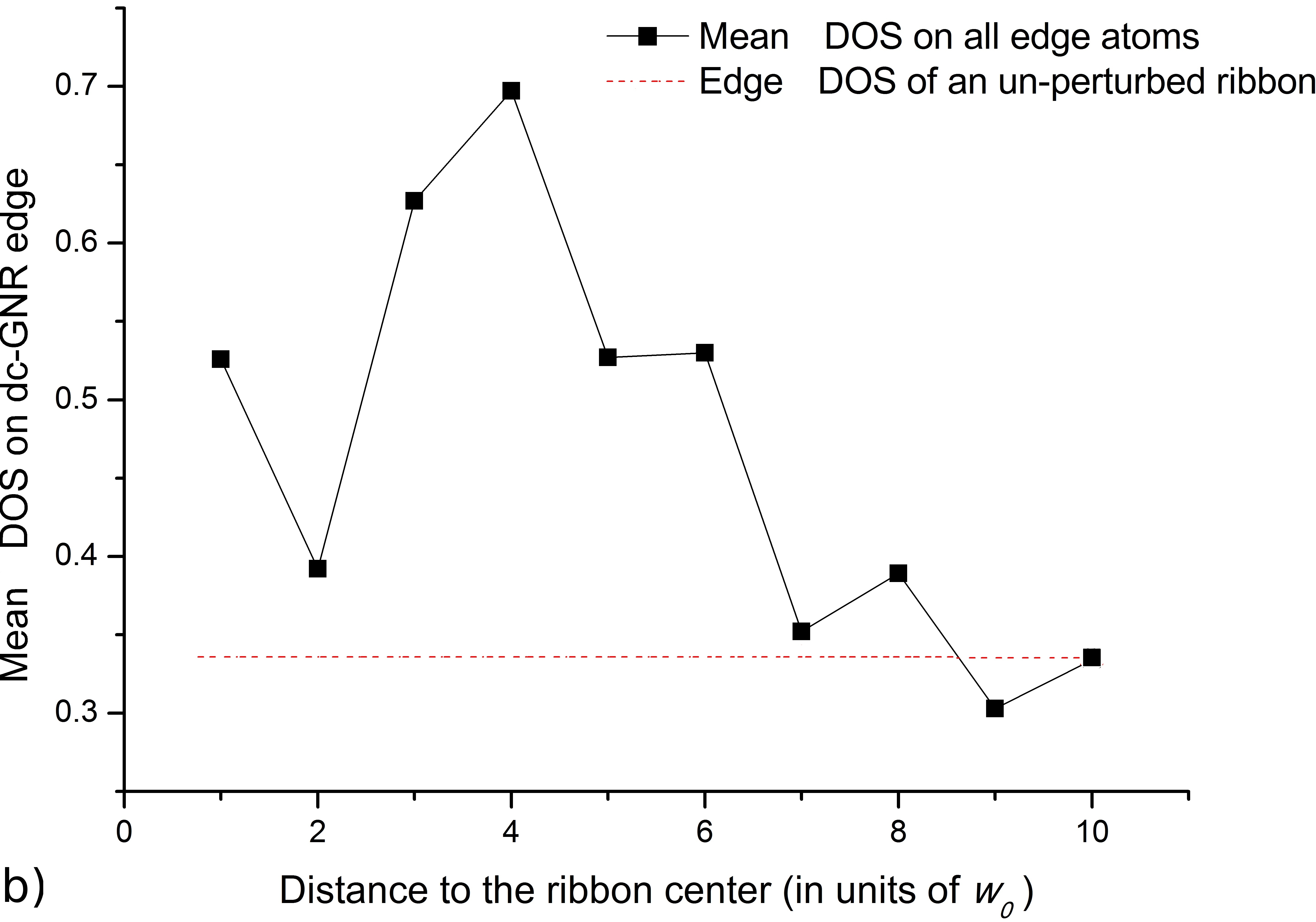}
\caption{TB (a) and DFT (b) calculated mean edge DOS as a function of the distance between the defect and the ribbon center for a $n=15$ ribbon (a) and $n=10$ (b).}
\label{fig5}
\end{figure}

\begin{figure}[t] 
\centering
\hspace{.8cm}
\includegraphics[width=7.5cm]{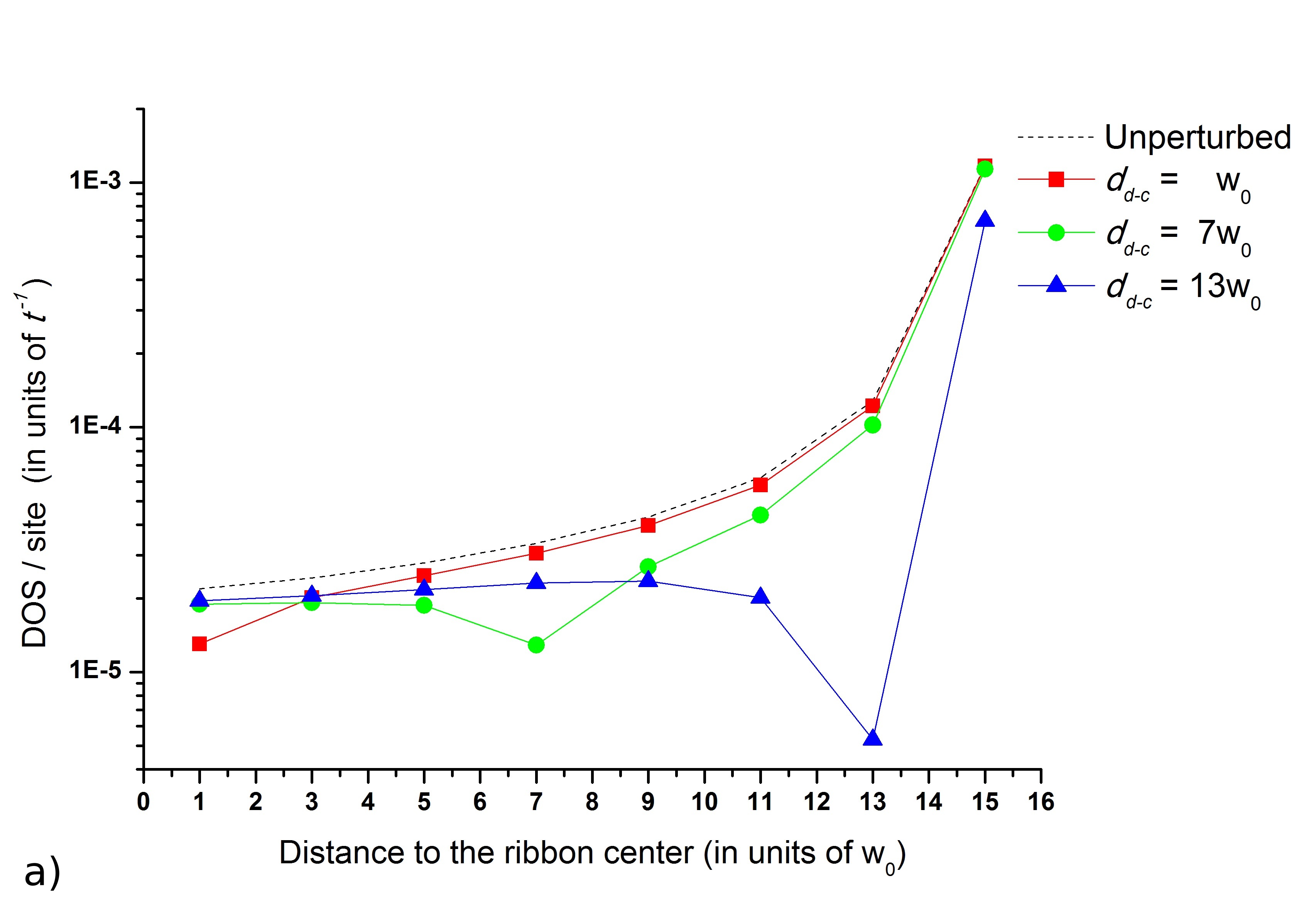}\\
\vspace{.5cm}
\includegraphics[width=8cm]{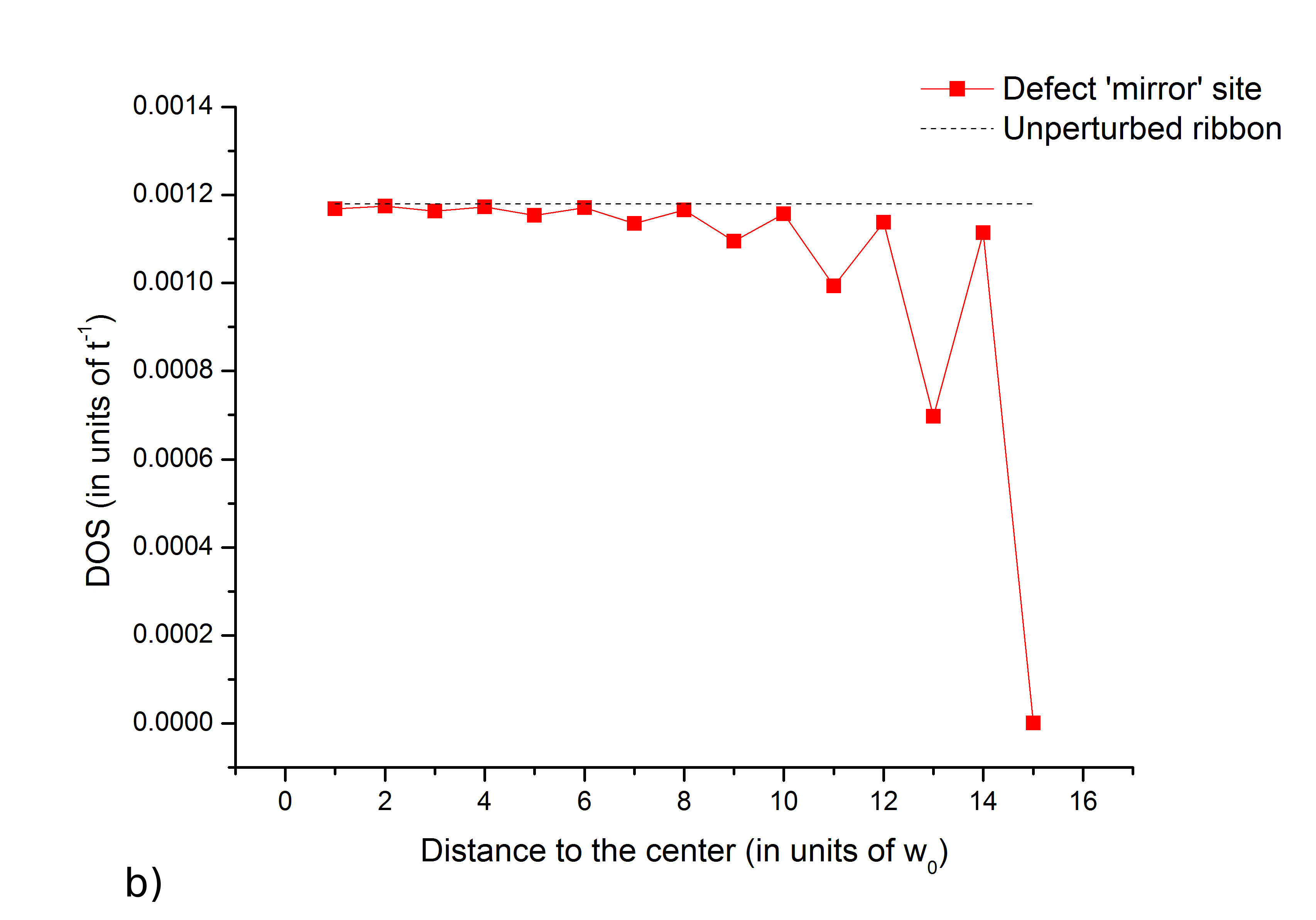}
\caption{a) LDOS along an axis perpendicular to the edges of the $n=15$  zz-GNR and passing trough the defect, for three positions of a $V = 10.0$ defect ($d_{d-c}$ is the distance between the defect and the ribbon center). b) The LDOS on the impurity mirror site as a function of the distance between the impurity and the edge.}
\label{fig6}
\end{figure}

\paragraph{The variation of the LDOS along the transversal direction}
\label{secd3}
In what follows we describe the dependence of the LDOS as a function of the distance from the edge, for different positions of the defect (see Fig.~\ref{fig6}). The LDOS along the transverse line defined in Fig.~\ref{fig1} that passes through the impurity is strongly perturbed, so that it exhibits a minimum on the impurity site.  Note that the LDOS on the impurity mirror site depends strongly on the impurity position, decreasing with the distance between the impurity and the edge (see Fig.~\ref{fig6} b). The odd/even oscillations in  Fig.~\ref{fig6} b) are probably the result of the difference in the definition of the impurity mirror site for different odd/even distances between the impurity and the edge (see Fig.~\ref{fig1}). 


\subsubsection{Two-dimensional behavior of the LDOS}

\begin{figure}[t]
\centering
\includegraphics[width=4cm]{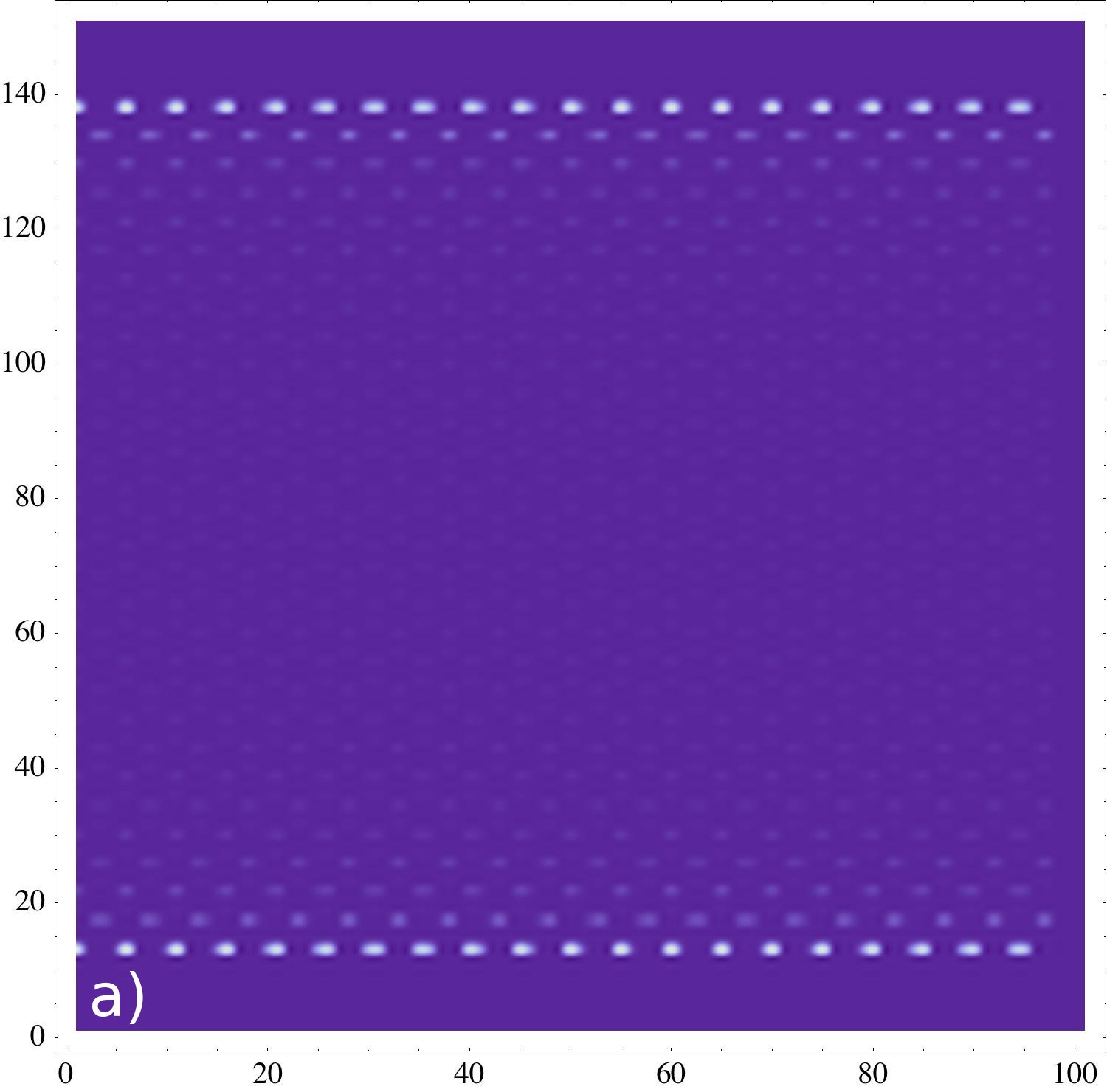}
\includegraphics[width=4cm]{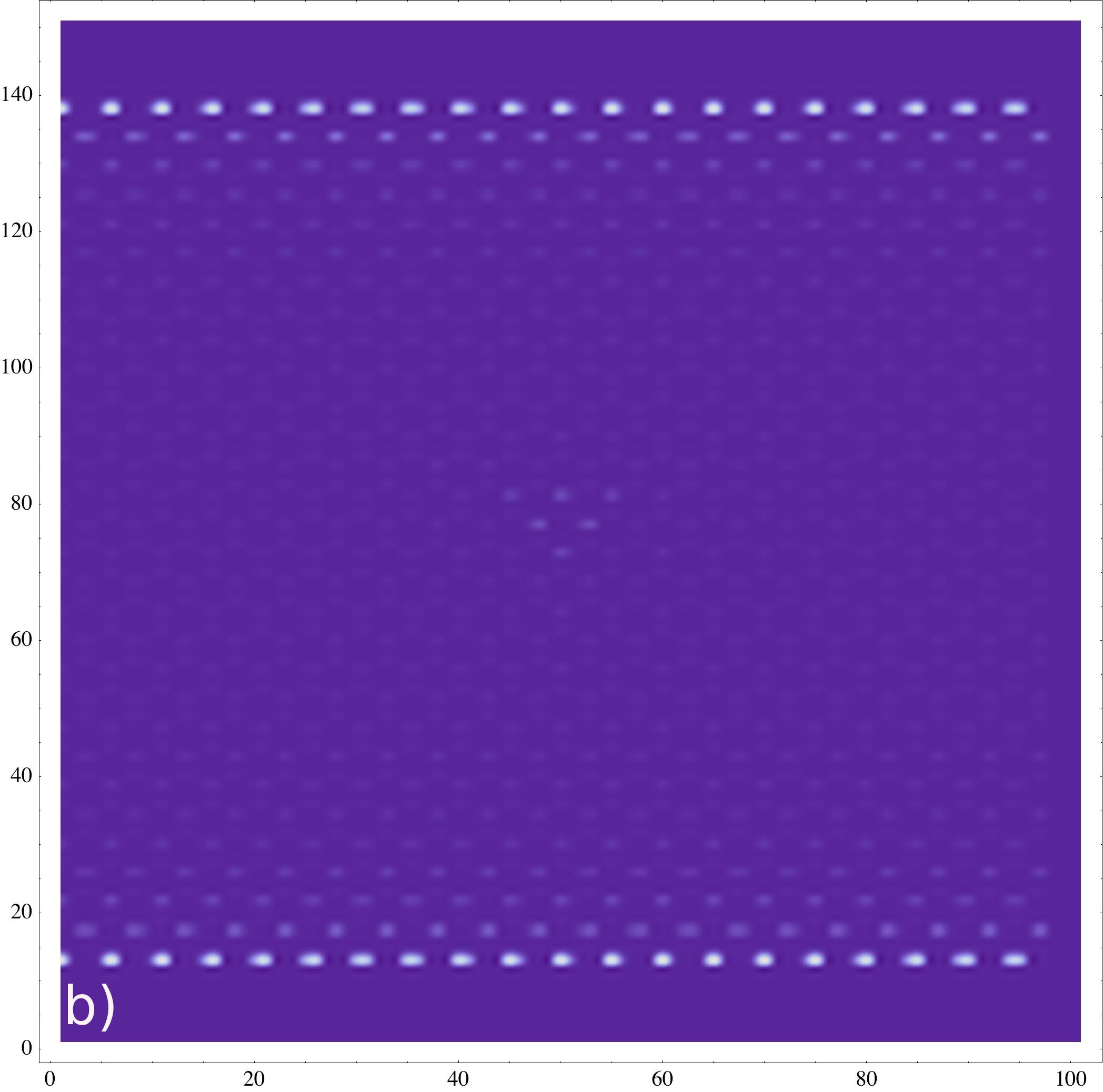}\\
\includegraphics[width=4cm]{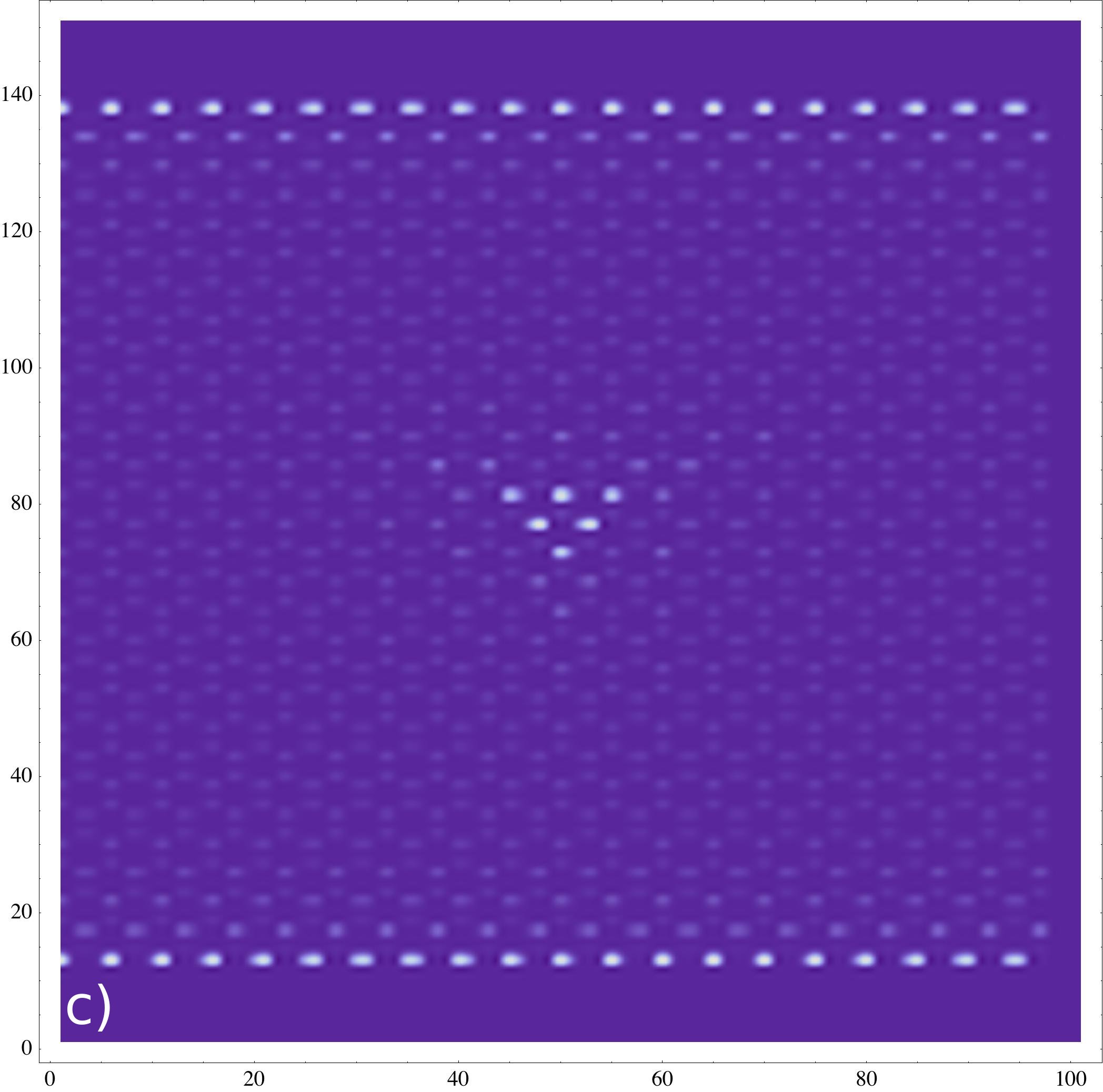}
\includegraphics[width=4cm]{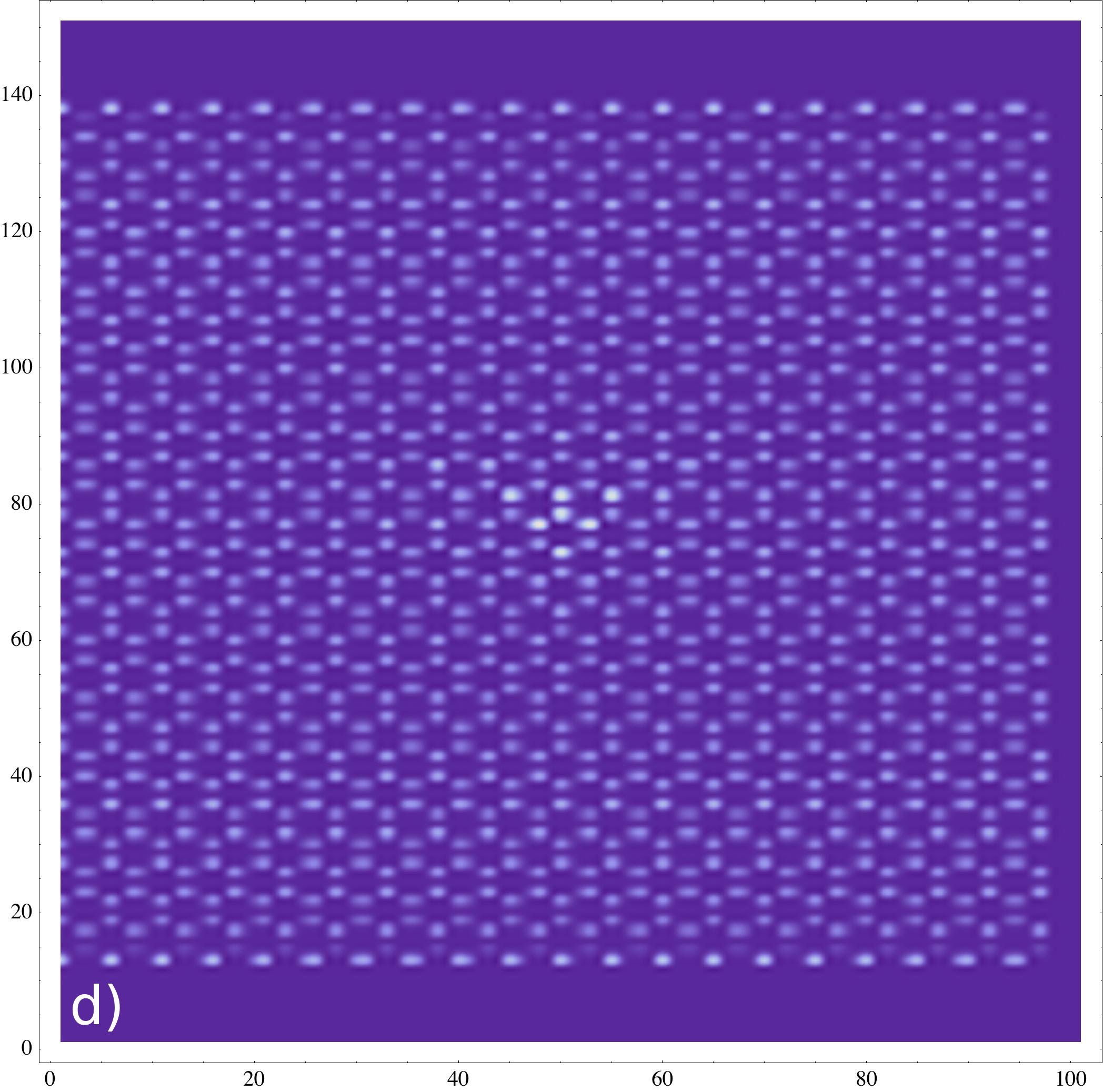}
\caption{TB calculated LDOS in $n=15$ (a) df-GNR ; (b) dc-GNR, for $E = 0$, for $V = 10.0$ ; (c) dc-GNR, for $E = -0.08$, $V = 10.0$ ; (d) dc-GNR,for $E = -0.38$, $V = 2.0$. }
\label{fig8}
\end{figure}

\begin{figure}[t]
\centering
\includegraphics[width=5cm]{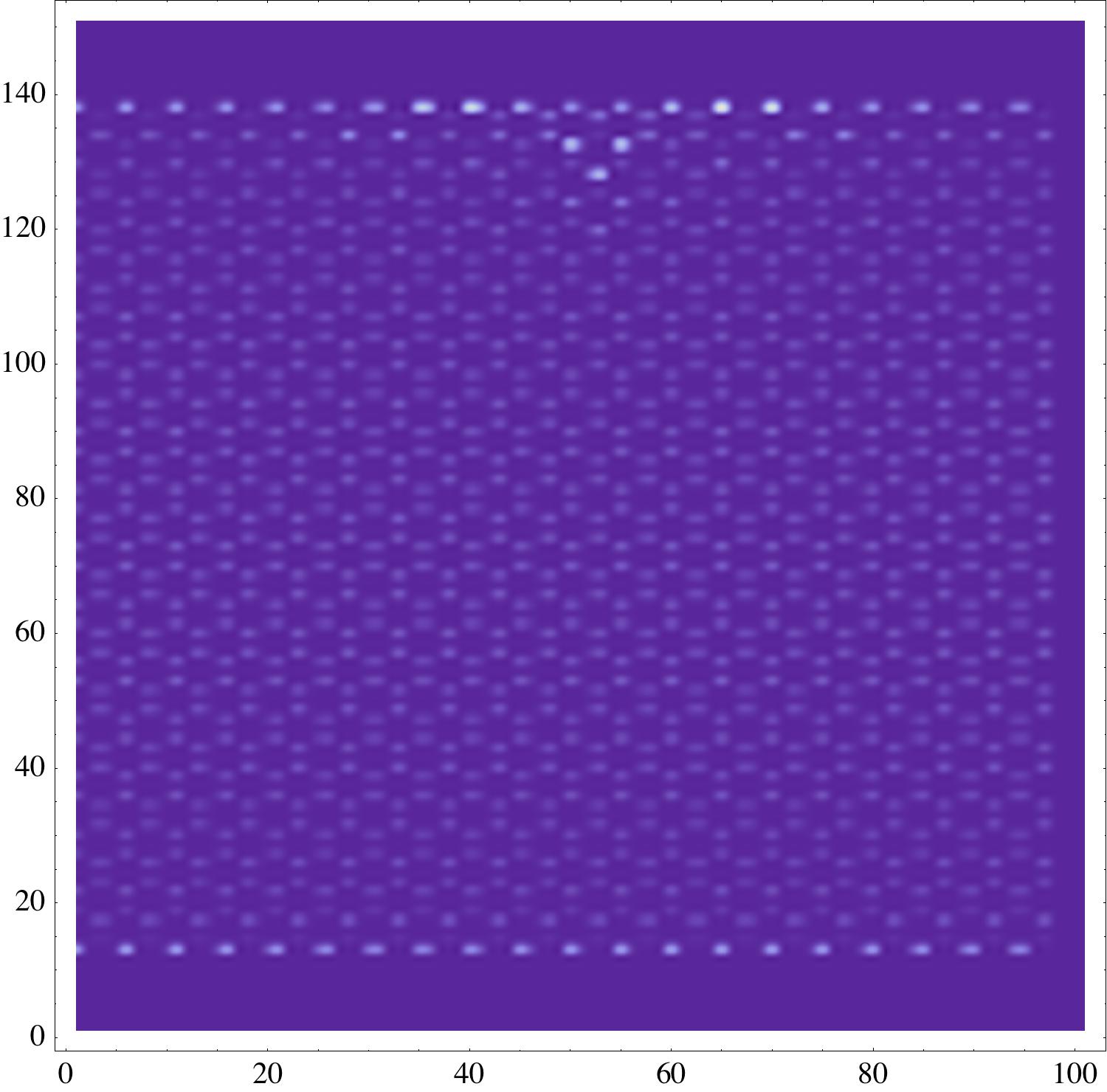}
\caption{The LDOS as a function of position when the impurity is approaching the edge. Note the reduction in intensity on the edge mirror site.}
\label{closeimp}
\end{figure}

We now focus on the two-dimensional position dependence of the LDOS. In Figs.~\ref{fig8},\ref{closeimp} we describe the behavior of the LDOS for a df-GNR as well as for two impurity positions and different values of the impurity potential. For a clean GNR one observes that the zero-energy DOS is confined to edge, and is constant along the longitudinal direction (see Fig. \ref{fig8}a) ). The intensity of the zero-energy LDOS is decaying drastically with the distance from the edge, such that it has a significant intensity only along the first few atomic rows parallel to the edge, while vanishing completely inside the df-GNR.  In the presence of a defect the zero-energy LDOS is non-zero both on the edges, as well as in the vicinity of the defect (see Fig.~\ref{fig8} b) for  $V = 10$). The LDOS amplitude close to the impurity is however maximal at a non-zero energy of  $E\approx 0.08$ for $V = 10$ and   $E \approx 0.34$ for $V = 2$, corresponding to the formation of the impurity state described in the previous section. Thus, for $V = 10$ (Fig.~\ref{fig8} c) and $E = 0.08$ the LDOS shows a maximum of intensity on the six atoms in the vicinity of the defect. However, for  $V = 2$, at energy $E\approx 0.34$ corresponding to the formation energy of the impurity state, the LDOS exhibits a  superposition between a bulk GNR state and the defect state  (see Fig.~\ref{fig8} d) .

In Fig.~\ref{closeimp} we plot the LDOS dependence on position, for an energy corresponding to the impurity state, when the impurity is close to the edge. Note the reduction in the intensity of the LDOS on the edge close to the impurity site, as well as the edge LDOS longitudinal oscillations.

\subsection{Density functional calculations}
We present here similar types of results as in the previous section, obtained this time via DFT calculations. Such calculations do not have access to the eigenstates of the system, but only to the LDOS. The defect considered here is a vacancy; in the tight-binding model this would correspond to an infinite onsite potential.

\begin{figure}[t]
\centering
\includegraphics[width=7cm]{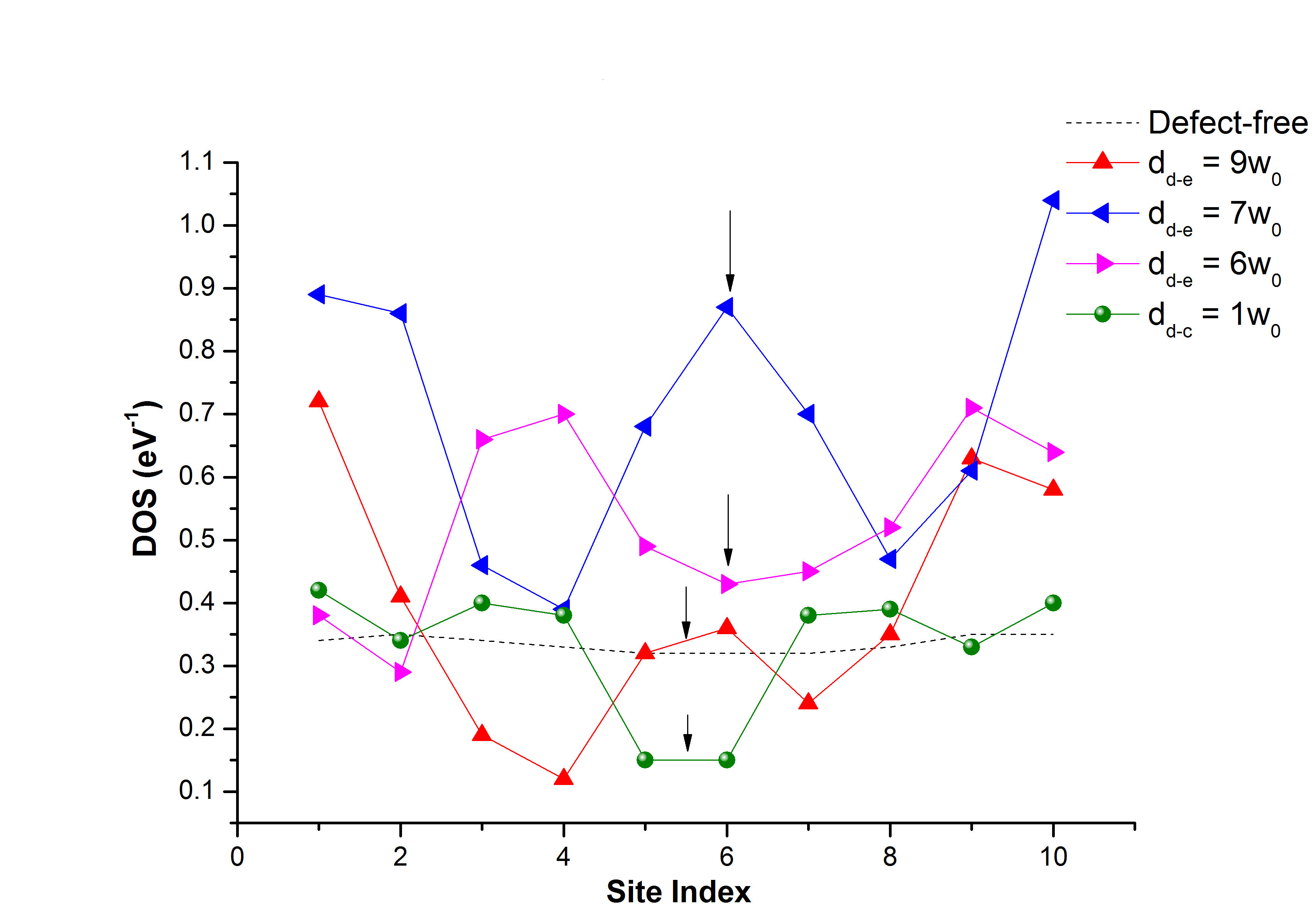}
\caption{The zero-energy edge DOS calculated via DFT for various defect-to-center distances, for a $n=10$ ribbon.}
\label{fig9}
\end{figure}

Similarly to the TB calculations, the DFT calculations show that a vacancy perturbs the LDOS on the edge. The shape of the edge perturbation changes rapidly with the distance to the edge, as  we describe in Fig.~\ref{fig9}. Thus, when the distance between the center of the GNR and the defect is an even/odd multiple of $w_0$, the induced edge DOS fluctuations have a maximum/minimum on the defect mirror site. This pattern breaks down when the impurity is very close to the edge, in which situation the mirror site always exhibits a minimum of DOS intensity. The irregularity of the observed modifications stems from the smaller size of the system, as well as from the calculation errors, inherently larger in the DFT calculations.

\begin{figure}[t]
\centering
\includegraphics[width=3cm]{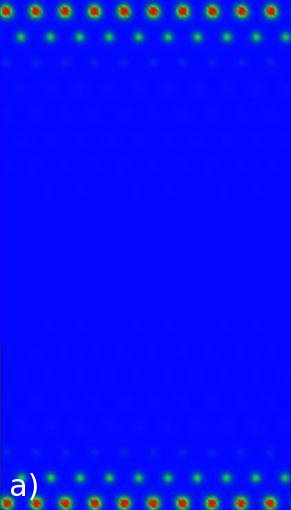}
\includegraphics[width=3cm]{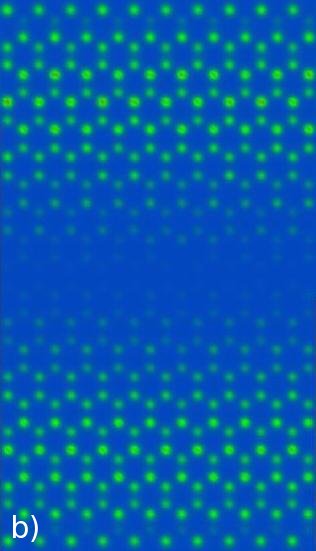}\\
\includegraphics[width=3cm]{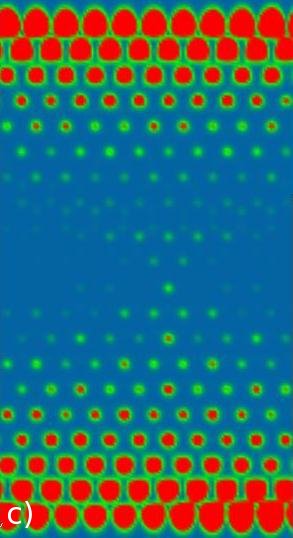}
\includegraphics[width=3cm]{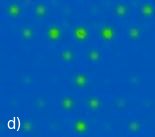}
\caption{DFT calculated LDOS plots in a zz-GNR $(n=10)$: (a) df-GNR ($E = 0.0 eV$); (b) bulk state ($E = - 1.0$);  (c) dc-GNR (vacancy state visible at $E = 0.0$); (d) zoom-in close to the defect  for the state described in c).}
\label{fig10}
\end{figure}

We note that the average DOS on the edge decreases with the distance between the impurity and the edge (see Fig.\ref{fig5} b), in a manner which is qualitatively similar to the the one predicted by the TB calculations.  When analyzing the two-dimensional LDOS profile obtained via the DFT methods (see Fig.~\ref{fig10}), we note that, similar to the TB results, the effect of the impurity consists mainly in the modification of the intensity on the six sites surrounding  the defect \cite{stmtheory1}. 


\subsection{Comparison between the infinite system and the finite size systems via the T-matrix approximation}

\begin{figure}[t]
\centering
\includegraphics[width=8cm]{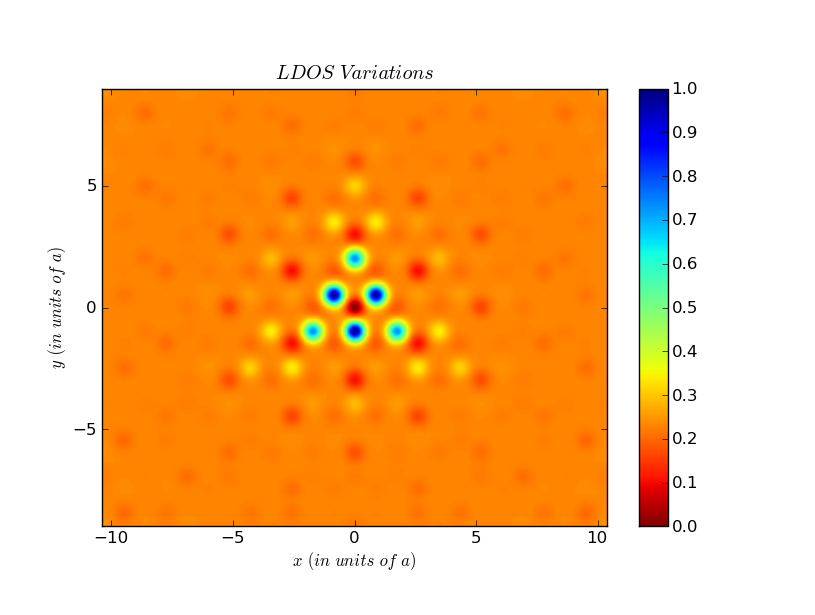}
\caption{Perturbation of the LDOS at the $w=0.15$ in the presence of a vacancy (infinite $V$) obtained using the T-matrix approximation.}
\label{figtm}
\end{figure}

To check the consistency of the DFT and TB calculations in describing the effects of the impurity on the LDOS we compare these results with those obtained for an infinite system via a T-matrix calculation \cite{stmtheory1}. The spatial dependence of the LDOS close to the impurity for an infinite grapheme sheet with a single impurity potential is presented in Fig.~\ref{figtm}. Note that, close to the impurity, the behavior of the LDOS is exactly the same as the one predicted by the TB and the DFT calculations. Note also that the long-wavelength oscillations that have been predicted by analytical calculations to decay as $1/r$ in a regular 2d system, and as $1/r^2$ in graphene \cite{stmtheory1}, decay too fast to be observed in either analytical T-matrix calculations, as well as in the TB and DFT calculations. We have checked that such oscillations are well captured and visible in the T-matrix analysis of a bilayer graphene.

\section{Conclusions}
We have studied the effect of a single impurity, modeled as an onsite potential, on the spectrum and LDOS of a zigzag GNR as a function of the impurity position and size of the impurity potential. We have found that, at zero energy, the amplitude of the DOS on the edge of the ribbon diminishes when the distance between the impurity and the edge decreases. Secondly we have noted oscillations along the ribbon edge in the zero-energy LDOS, whose amplitude increases when the impurity is approaching the edge. Thirdly, we have found that the eigenvalues and eigenfunctions of the system are modified, such that the lowest-energy wavefunctions exhibit a change in their longitudinal periodicity, while approximatively half of the finite-energy eigenvalues are shifted, and the corresponding eigenfunctions hybridize with the impurity state. The largest hybridization and eigenvalue shift occur for a state which we denote the impurity state, and which is localized in the vicinity of the defect. 

Our analysis indicates that the finite-size disordered GNR is a complex systems,  which exhibits extended size perturbations due to the interference effects between the finite-size effects and the disorder. Thus, a perturbative analysis of the disorder effects is not sufficient to describe the physics of such systems, but non-perturbative approaches such as DFT and tight-binding methods are necessary.

\acknowledgments We would like to thank Gilles Montambaux for useful discussions. This work has been supported by the ANR project NANOSIM GRAPHENE under Grant No. ANR-09- NANO-016, and by the FP7 ERC Starting Independent Researcher Grant NANO-GRAPHENE 256965.

\end{document}